\documentclass[iop]{emulateapj}

\usepackage{placeins}

\usepackage{amsmath}  

\usepackage{color}
\definecolor{linkcolor}{rgb}{0.85, 0, 0.5}
\usepackage[colorlinks=true,linkcolor=linkcolor,anchorcolor=blue,citecolor=blue,urlcolor=blue, draft=false]{hyperref}

\newcommand{\LCDM}{$\Lambda$CDM}

\renewcommand{\~}{$\sim$}

\newcommand{\um}{$\mu$m}

\newcommand{\sig}{$\sigma$}
\newcommand{\Lya}{Lyman-$\alpha$}

\newcommand{\super}[1]{$^{\rm #1}$}

\newcommand{\supa}{$^{\rm a}$}
\newcommand{\supb}{$^{\rm b}$}
\newcommand{\supc}{$^{\rm c}$}
\newcommand{\supd}{$^{\rm d}$}
\newcommand{\supe}{$^{\rm e}$}
\newcommand{\supf}{$^{\rm f}$}
\newcommand{\supg}{$^{\rm g}$}
\newcommand{\suph}{$^{\rm h}$}

\newcommand{\squared}{$^2$}
\newcommand{\cubed}{$^3$}

\newcommand{\sqarcmin}{arcmin\squared}

\newcommand{\chisq}{$\chi^2$}

\newcommand{\per}{$^{-1}$}
\newcommand{\inv}{\per}
\newcommand{\Mstar}{$M^*$}

\newcommand{\phistar}{$\phi^*$}

\newcommand{\Om}{\Omega_m}
\newcommand{\OL}{\Omega_\Lambda}

\newcommand{\HST}{{\em HST}}
\newcommand{\SST}{{\em SST}}
\newcommand{\Hubble}{{\em Hubble}}
\newcommand{\Spitzer}{{\em Spitzer}}

\newcommand{\citepeg}[1]{\citep[e.g.,][]{#1}}

\newcommand{\range}[2]{\! \left[ _{#1} ^{#2} \right] \!}  

\shortauthors{Coe, Bradley, \& Zitrin 2014}

\slugcomment{Submitted to the Astrophysical Journal, \today}

\usepackage{datetime}

\begin{document}


\title{Frontier Fields: High-Redshift Predictions And Early Results}
\shorttitle{Frontier Fields: High-Redshift Predictions And Early Results}
\hypersetup{pdfauthor={Dan Coe}, pdftitle={Frontier Fields: High-Redshift Predictions And Early Results}}



\author{Dan Coe\altaffilmark{1*}}
\author{Larry Bradley\altaffilmark{1}}
\author{Adi Zitrin\altaffilmark{2,3}}

\altaffiltext{*}{\email{DCoe@STScI.edu}}
\altaffiltext{1}{Space Telescope Science Institute, Baltimore, MD, USA}
\altaffiltext{2}{California Institute of Technology, Pasadena, CA, USA}
\altaffiltext{3}{Hubble Fellow}



\begin{abstract}

The Frontier Fields program is obtaining deep \Hubble\ and \Spitzer\ Space Telescope images 
of new ``blank'' fields and nearby fields gravitationally lensed by massive galaxy clusters.  
The \Hubble\ images of the lensed fields are revealing nJy sources (AB mag $>$ 31), 
the faintest galaxies yet observed.  
In this paper, we present high-redshift ($z > 6$) number count predictions for the full program
and candidates in three of the first \Hubble\ Frontier Fields images.
The full program will transform our understanding of galaxy evolution in the first 600 million years ($z > 9$).  
Where previous programs yielded perhaps a dozen $z > 9$ candidates, the Frontier Fields may yield \~70 (\~6 per field).  
We base this estimate on an extrapolation of luminosity functions observed between $4 < z < 8$
and gravitational lensing models submitted by the community.  
However, in the first two deep infrared Hubble images obtained to date, 
we find $z \sim 8$ candidates but no strong candidates at $z > 9$. 
This might suggest a deficit of faint $z > 9$ galaxies as also reported in the Ultra Deep Field
(even while excesses of brighter $z > 9$ galaxies were reported in shallower fields). 
At these redshifts, cosmic variance (field-to-field variation) is expected to be significant (greater than $\pm 50$\%) 
and include clustering of early galaxies formed in overdensities. 
The full Frontier Fields program will significantly mitigate this uncertainty 
by observing six independent sightlines each with a lensing cluster and nearby blank field.

\end{abstract}


\keywords{early universe --
galaxies: high redshift --
cosmology: dark ages, reionization, first stars --
galaxies: evolution --
gravitational lensing: strong --
galaxies: clusters
}



\section{Introduction}
\label{sec:intro}

Discovering the first galaxies \citep{BrommYoshida11, Dunlop13book}
and determining their role in reionizing the universe \citep{Robertson13}
are frontier research areas of extragalactic astronomy.
The {\em Hubble} and {\em Spitzer Space Telescopes} (\HST\ and \SST)
are steadily discovering more distant high-redshift galaxies at earlier times
with the Wide Field Camera 3 infrared channel (WFC3/IR; \citealt{WFC3}) and Infrared Array Camera (IRAC; \citealt{IRAC}), respectively.
%
Gravitational lensing by galaxy clusters \citep{KneibNatarajan11}
has helped boost the power of these telescopes
to observe faint sources \citepeg{Bradley08},
as in the
Cluster Lensing And Supernova survey with Hubble (CLASH; \citealt{CLASH}).
Prior to October 2013 (including all of CLASH), 
the deepest WFC3/IR observations of any strong lensing cluster were 7 orbits
(divided among the WFC3/IR filters).
Still, these have complemented ``blank'' field programs
such as the Ultra Deep Field (UDF; \citealt{Illingworth13, Ellis13}),
the Cosmic Assembly Near-Infrared Deep Extragalactic Legacy Survey
(CANDELS; \citealt{Grogin11, Koekemoer11}),
the WFC3/IR Early Release Science (ERS) observations \citep{Windhorst11},
and the \HST\ pure-parallel programs BoRG \citep{Trenti12} and HIPPIES \citep{Yan11}
in searching for the most distant galaxies yet known.

CLASH observations yielded a strongly lensed 
$z \sim 10.8$ candidate
observed \~420 Myr after the Big Bang
\citep{Coe13}.
A handful of candidates have been reported at $z \sim 9$ -- 10 
from CLASH \citep{Zheng12N, Bouwens12c},
the UDF \citepeg{Ellis13, Oesch13}, 
and CANDELS \citep{Oesch14}.
A more sizable sample of \~100 candidates has been discovered at $z \sim 8$
from CLASH \citep{Bradley13}
and blank fields \citepeg{Finkelstein12b, Schmidt14, Bouwens14}.
%
Ground-based observatories have contributed spectroscopic confirmations out to $z = 7.51$ \citep{Finkelstein13}
or 7.62 \citep{Schenker14}
plus additional high-redshift candidates confirmed out to $z = 7.21$ \citep{Shibuya12, Ono12}.

From these early times ($z \sim 8$),
the average cosmic star formation rate density (SFRD) steadily increased to its peak at $z \sim 2$
before declining to its modern day value as gas supplies dwindled
(see \citealt{MadauDickinson14} for a recent review).
%
At earlier times ($z > 8$),
a similarly steady evolution in SFRD 
may have been required to yield the galaxy stellar masses observed at $z \sim 8$ and later
\citep{BehrooziSilk14}.
But some observations suggest a more rapid buildup
at earlier times ($z > 8$)
than observed at $2 < z < 8$.
 \cite{Bouwens11N} and \cite{Oesch12a, Oesch13} report too few $z \sim 10$ candidates
at faint magnitudes (\~30 AB) in the UDF and surrounding fields. 
At brighter magnitudes (\~26--27 AB),
\cite{Oesch14} find an excess of 
$z \sim 9$ -- 10 candidates
in CANDELS observations of GOODS-N and GOODS-S,
where no such ``bright'' candidates were expected
(see also \citealt{Bouwens14}).

Improved statistics on $z \gtrsim 9$ luminosity functions
are forthcoming
the Frontier Fields program.
In October 2013, this program began obtaining deep \Hubble\ and \Spitzer\ imaging 
of six galaxy clusters and six blank fields in parallel.
The \Hubble\ images of the lensed fields are now our deepest views of the universe to date,
probing nJy sources (intrinsic AB magnitudes $>$ 31).
The WFC3/IR images (70 orbits) are 10 times deeper than any previous strong lensing cluster WFC3/IR imaging.
Improved constraints on faint number counts (which rise steeply with magnitude at high redshifts;
e.g., \citealt{Bradley12b})
will further constrain the ability of faint galaxies to reionize the early universe \citepeg{Bouwens12a}.
And where the Ultra Deep Field may, by chance, be overdense or underdense at a given redshift,
the six Frontier Fields sightlines significantly reduce uncertainties due to cosmic variance \citepeg{TrentiStiavelli08, Bouwens14}.

To date, the Frontier Fields program has obtained half of the \Hubble\ observations of four fields:
the galaxy cluster Abell 2744, 
its blank parallel field,
cluster MACSJ0416.1-2403,
and its blank parallel field.
Previous works presented high-redshift candidates lensed by Abell 2744,
based on the deep infrared \Hubble\ imaging and the first half of deep \Spitzer\ imaging
\citep{Atek14, Laporte14, Zheng14}.

In this paper, we identify new high-redshift candidates 
in the field lensed by MACSJ0416.1-2403 and its blank parallel field.
%
%
We also present, for the first time, predictions of lensed high-redshift number counts from the full Frontier Fields program
based on public lensing models of the clusters submitted by multiple teams.

Below we briefly describe the Frontier Fields program (\S\ref{sec:FF})
and the gravitational lensing models submitted by the community (\S\ref{sec:lensmodels}).
We use these lens models along with extrapolated luminosity functions
to predict high-redshift galaxy yields from the program (\S\ref{sec:predictions}).
Based on the \Hubble\ Frontier Fields observations to date (\S\ref{sec:obs}),
we identify high-redshift candidates (\S\ref{sec:highz})
and conclude by discussing our results (\S\ref{sec:conclusions}).

We assume a flat concordance \LCDM\ cosmology
with $h = 0.7$, $\Om = 0.3$, and $\OL = 0.7$,
where $H_0 = 100 \, h$ km s\inv\ Mpc\inv.

\begin{figure}
\centerline{
\includegraphics[width = \columnwidth]{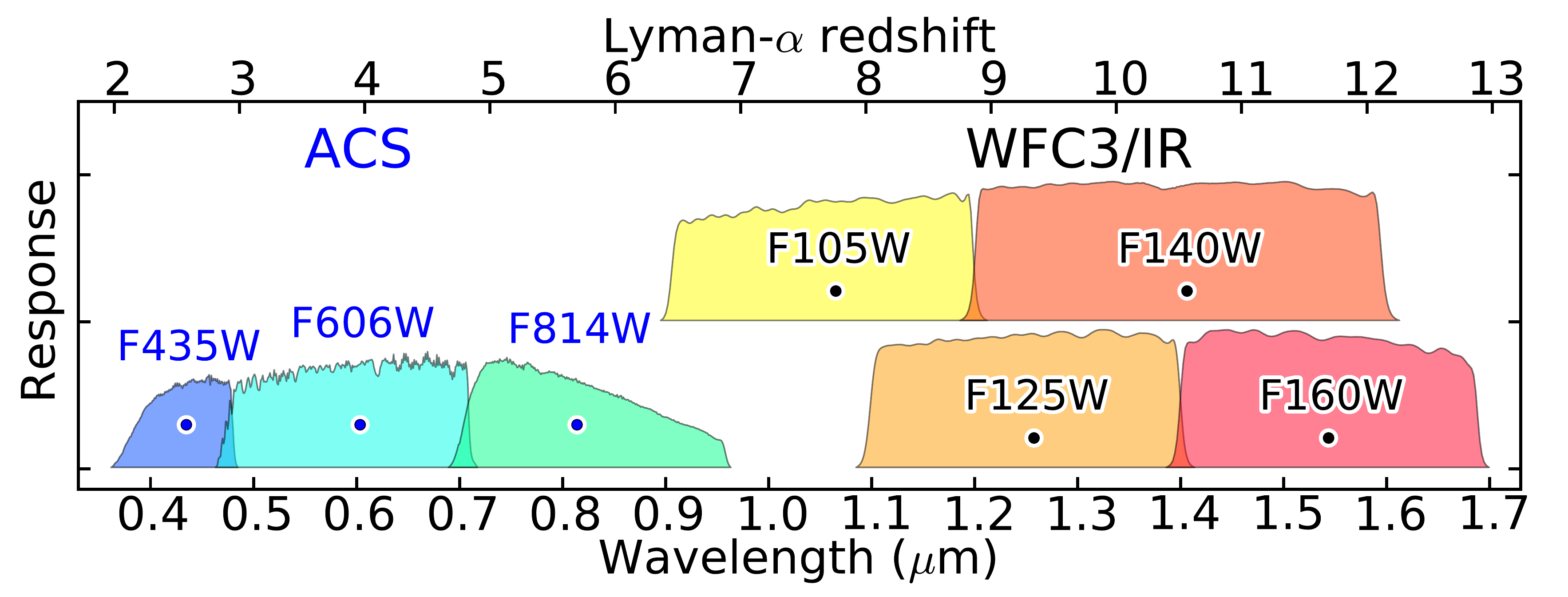}
}
\caption{
\label{fig:filters}
The Frontier Fields are being observed with these seven \Hubble\ ACS and WFC3/IR filters.
Response curves are plotted versus wavelength ($\lambda$)
with the corresponding \Lya\ redshift ($z$) given along the top axis ($\lambda = 0.1216\mu$m $(1+z)$).
F105W and F140W are offset vertically for clarity.
Dots mark the effective ``pivot'' wavelengths \citep{TokunagaVacca05} of the filters.
}
\end{figure}

\begin{deluxetable}{lllccr}
\tablecaption{\label{tab:filters}Exposure Times and Expected Depths}
\tablewidth{\columnwidth}
\tablehead{
\colhead{}&
\colhead{}&
\colhead{$\lambda$\supa}&
\colhead{}&
\colhead{Depth\supb}&
\colhead{Sens.\supc}
\\
\colhead{Camera}&
\colhead{Filter}&
\colhead{(\um)}&
\colhead{Exposure}&
\colhead{(AB)}&
\colhead{(nJy)}
}
\startdata
\HST\ ACS/WFC & F435W & 0.43 & 18 orbits & 28.8 & 2.2\\
\HST\ ACS/WFC & F606W & 0.59 & 10 orbits & 28.8 & 2.2\\
\HST\ ACS/WFC & F814W & 0.81 & 42 orbits & 29.1 & 1.7\\
\HST\ WFC3/IR & F105W & 1.06 & 24 orbits & 28.9 & 2.0\\
\HST\ WFC3/IR & F125W & 1.25 & 12 orbits & 28.6 & 2.6\\
\HST\ WFC3/IR & F140W & 1.39 & 10 orbits & 28.6 & 2.6\\
\HST\ WFC3/IR & F160W & 1.54 & 24 orbits & 28.7 & 2.4\\
\SST\ IRAC & ch1 & 3.6 & 50 hours & 26.6 & 16.6\\
\SST\ IRAC & ch2 & 4.5 & 50 hours & 26.0 & 28.9
\enddata
\tablenotetext{1}{Effective ``pivot'' wavelength \citep{TokunagaVacca05}}
\tablenotetext{2}{5\sig\ point source AB magnitude limit (within a 0.4\arcsec\ diameter aperture for \HST)}
\tablenotetext{3}{1\sig\ point source sensitivity (nJy) within the same apertures}
\end{deluxetable}


\section{Frontier Fields}
\label{sec:FF}

The Frontier Fields\footnote{
\href{http://www.stsci.edu/hst/campaigns/frontier-fields/}
{http://www.stsci.edu/hst/campaigns/frontier-fields/}}
(PI Lotz) is a \Hubble\ and \Spitzer\ Space Telescope program
to obtain deep images of six strongly lensing clusters and six nearby ``blank'' fields
6\arcmin\ from the cluster cores.
After deliberating over white papers submitted by the community,
the Hubble Deep Fields Initiative (HDFI) science working group unanimously recommended this program\footnote{
\href{http://www.stsci.edu/hst/campaigns/frontier-fields/HDFI\_SWGReport2012.pdf}
{http://www.stsci.edu/hst/campaigns/frontier-fields/\\
HDFI\_SWGReport2012.pdf}}
to the directors of the
Space Telescope Science Institute (STScI)
and the
Spitzer Science Center (SSC).
The directors approved observations of the first four pairs of clusters and blank fields
with approval for the final two pairs pending an early interim review.
The complete \Hubble\ program would span three years (Cycles 21--23).

The six galaxy clusters to be observed are, in order,
Abell 2744 ($z = 0.308$), 
MACSJ0416.1-2403 ($z = 0.396$), 
MACSJ0717.5+3745 ($z = 0.545$), 
MACSJ1149.5+2223 ($z = 0.543$), 
Abell S1063 ($z = 0.348$), 
and 
Abell 370 ($z = 0.375$).
These clusters
were discovered by \cite{Abell58}, \cite{Abell89},
and the Massive Cluster Survey (MACS; \citealt{Ebeling07}, \citealt{MannEbeling12}).

\Hubble\ is imaging each cluster and ``blank'' field for 140 orbits
with the Advanced Camera for Surveys (ACS) and the Wide Field Camera 3 (WFC3) IR channel
working in parallel with seven filters spanning 0.4--1.7\um.
The 5\sig\ detection limit for point sources is roughly 29th magnitude AB in each filter (Table \ref{tab:filters}).
\Spitzer\ is imaging each field for 100 hours total
with the Infrared Array Camera (IRAC)
in channels 1 and 2 spanning 3--5\um.\footnote{
\href{http://ssc.spitzer.caltech.edu/warmmission/scheduling/approvedprograms/ddt/frontier/}
{http://ssc.spitzer.caltech.edu/warmmission/scheduling/\\approvedprograms/ddt/frontier/}}
The 50-hour 3.6\um\ and 4.5\um\ images are expected to reach AB mag 26.6 and 26.0, respectively (5\sig\ point source).

\begin{figure*}
\centerline{
\includegraphics[width = 0.8\textwidth]{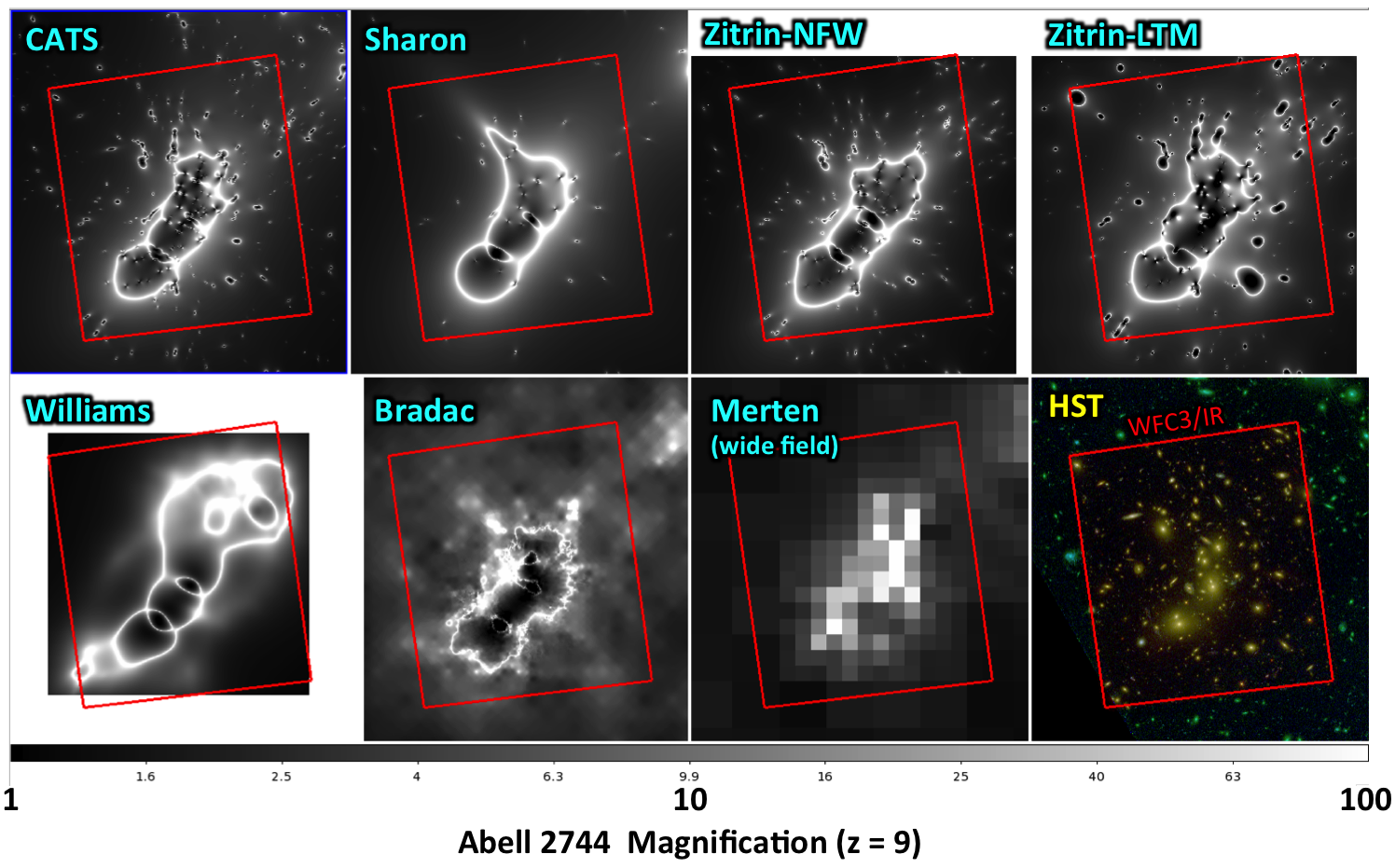}
}
\caption{
\label{fig:A2744models}
Magnification maps (log grayscale) for $z = 9$ galaxies lensed by Abell 2744 
according to all seven submitted gravitational lensing models.
The Frontier Fields WFC3/IR FOV ($136''  \times 123''$) is outlined in red.
At bottom right is a color HST image 
(produced with Trilogy; \citealt{Coe12})
showing the Frontier Fields WFC3/IR observations (red channel)
within the prior ACS observations (blue-green).  North is up, and East is left.
}
\end{figure*}

\section{Gravitational Lensing Models}
\label{sec:lensmodels}

Based on further recommendations from the HDFI science working group,
STScI commissioned five teams to produce gravitational lensing models for all six Frontier Fields clusters
to be made available to the community.
All teams shared the latest observational constraints,
including positions and redshifts of galaxies strongly lensed to form multiple images,
before working independently on the lensing models.
Some teams also included weak lensing constraints in their modeling.
For each cluster, the teams altogether submitted seven or eight models they deemed ``best''
as well as ranges of models yielding the uncertainties for each method.
All of these models are available to the public along with tools to facilitate their use.\footnote{
\href{http://archive.stsci.edu/prepds/frontier/lensmodels/}{http://archive.stsci.edu/prepds/frontier/lensmodels/}}

In Figure \ref{fig:A2744models}, we show high-redshift ($z = 9$) magnification maps from the seven ``best'' models submitted for Abell 2744.
The various methodologies are summarized online\footnote{
\href{http://www.stsci.edu/hst/campaigns/frontier-fields/Lensing-Models}{http://www.stsci.edu/hst/campaigns/frontier-fields/Lensing-Models}}
and described in more detail in each team's README files.
Here we briefly describe the various models.

The ``CATS'' (Clusters As Telescopes) team 
(including Co-PIs J.-P.~Kneib and P.~Natarajan; Admin PI H.~Ebeling; modelers J.~Richard and M.~Limousin)
and the K.~Sharon (PI) group including modeler T.~Johnson both use Lenstool \citep{Kneib93, Jullo07, Jauzac12}.
They model each cluster's mass as one or more large halos 
plus smaller subhalos at the positions of cluster galaxies identified in the images.
Each halo and subhalo is modeled as a 
pseudo-isothermal elliptical mass distribution (PIEMD; \citealt{KassiolaKovner93}).
In this work we use the ``version 2'' models from the Sharon group
and version 1 from all others.

A.~Zitrin (Co-PI with J.~Merten) independently implemented a similar parameterization for the ``Zitrin-NFW'' models
with PIEMDs for the galaxy halos and elliptical NFW \citep{NFW96} mass distributions for the cluster halos 
(as in \citealt{Zitrin13MACS0416}).
He also submitted ``LTM'' models \citep{Broadhurst05, Zitrin09a} in which light traces mass more completely:
in addition to cluster galaxies tracing the positions of subhalos,
the cluster halo is modeled as a smoothed version of the cluster galaxy light.
For these models, A.~Zitrin used two smoothing kernels: polynomial splines (``Zitrin-LTM'') and Gaussians (``Zitrin-LTM-Gauss'').
For each cluster, he submitted models based on two or three of these methods (NFW, LTM, and LTM-Gauss).\footnote{
Zitrin-LTM is available for all six clusters.
Zitrin-LTM-Gauss is available for all but Abell 2744,
and Zitrin-NFW for all but MACSJ0717.5+3745 and MACSJ1149.5+2223.}

Others submitted models which made no assumptions about light tracing mass,
forgoing that useful prior in favor of a broader range of allowed solutions.
The M.~Bradac (PI) team including A.~Hoag used the method from \cite{Bradac05, Bradac09}.
The L.~Williams (PI) team used Grale \citep{Liesenborgs06, Liesenborgs07}.
J.~Merten (Co-PI with A.~Zitrin) used SaWLens \citep{Merten09, Merten11}
to produce lower resolution models of wider areas extending out to include the parallel ``blank'' fields
based on observed weak lensing.

\subsection{Lens Model Products}

%
Pre-computed magnification maps are available online at four redshifts ($z = 1, 2, 4, 9$)
along with maps of the mass surface density $\kappa$ and weak lensing shear $\gamma$,
from which we derive magnification maps at other redshifts using the scripts provided online.
Magnifications $\mu$ are calculated as
$1 / \mu = | (1 - \kappa)^2 - \gamma^2 |$
after $\kappa$ and $\gamma$ have been scaled to the desired redshift.
%
Both $\kappa$ and $\gamma$ are proportional to $D_{LS} / D_{S}$,
a ratio of angular-diameter distances from lens to source 
$D_{LS} = D_A(z_L,z_S)$
and observer to source
$D_S = D_A(0,z_S)$.
%
For a flat universe ($\Omega = \Omega_m + \Omega_\Lambda = 1$), 
angular-diameter distances are calculated as follows 
(\citealt{Fukugita92}, filled beam approximation; see also \citealt{Hoggcosmo}
and online calculators such as 
iCosmos\footnote{\href{http://www.icosmos.co.uk}{http://www.icosmos.co.uk}}):
\begin{eqnarray}
  D_A(z_1,z_2) & = & \frac{c}{1 + z_2} \int_{z_1}^{z_2} \frac{dz\prime}{H(z\prime)}\\
  & = & D_A(0, z_2) - \left( \frac{1+z1}{1+z2} \right) D_A(0, z_1),
\end{eqnarray}
where (again for a flat universe) the Hubble parameter varies with redshift as
$H(z) = H_o \sqrt{\Omega_m (1 + z)^3 + \Omega_\Lambda}$.

All lens modelers who assumed light traces mass to some degree (CATS, Sharon, Zitrin)
also submitted models of the lensing deflection fields $\vec \alpha$.
These enable us to study the delensed source plane as predicted by each model.
The deflections are also proportional to $D_{LS} / D_{S}$;
the online maps must be scaled by this factor.

\begin{figure*}
\centerline{
\includegraphics[width = 0.8\textwidth]{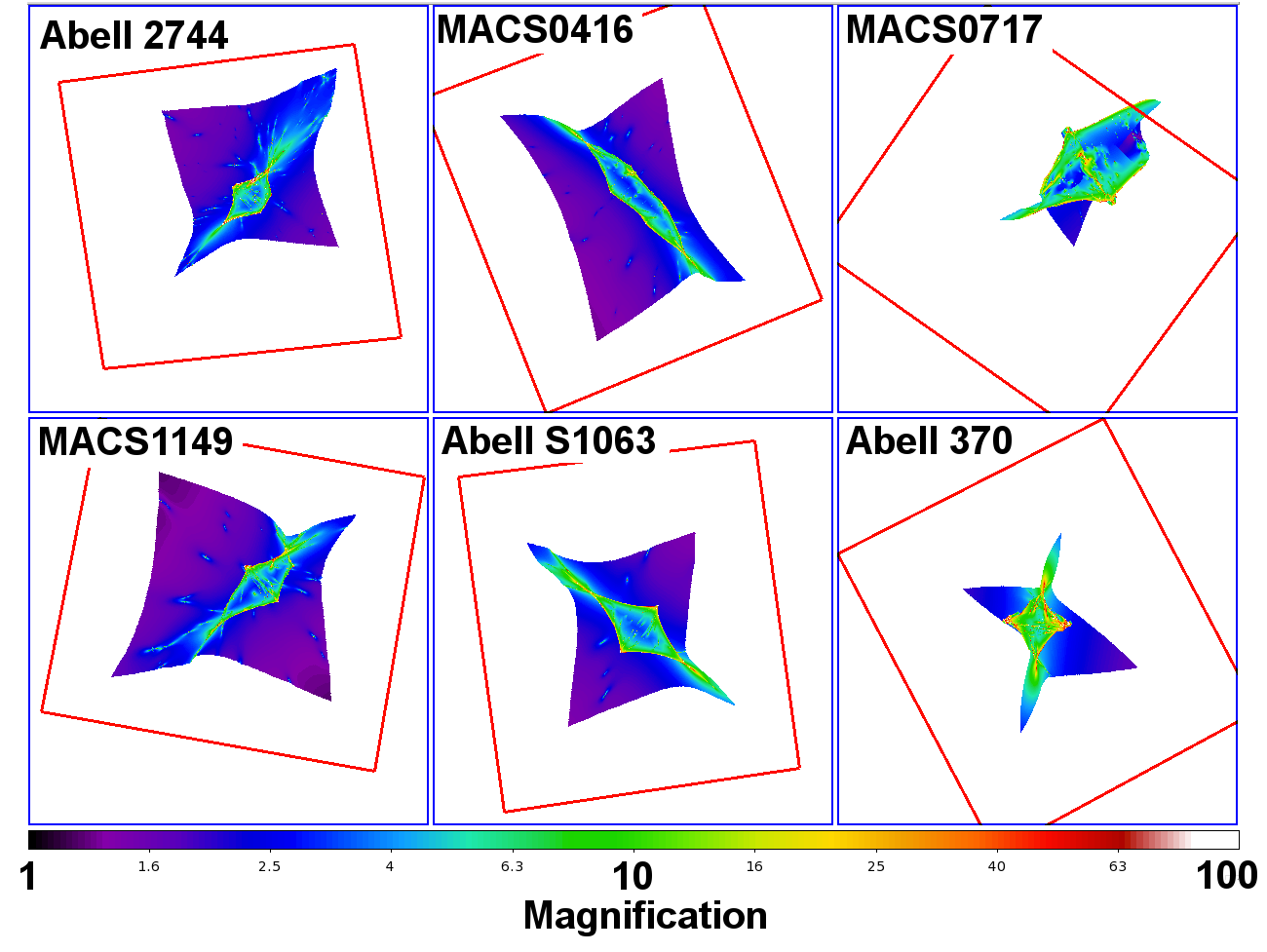}
}
\caption{
\label{fig:sourceplanes}
Examples of delensed $z = 9$ magnifications maps (according to the CATS models) 
of the WFC3/IR fields (red outlines $136'' \times 123''$)
to be imaged for each Frontier Fields cluster.  
In each panel, north is up and east is left.
}
\end{figure*}

\begin{figure}
\centerline{
\includegraphics[width = 0.925\columnwidth]{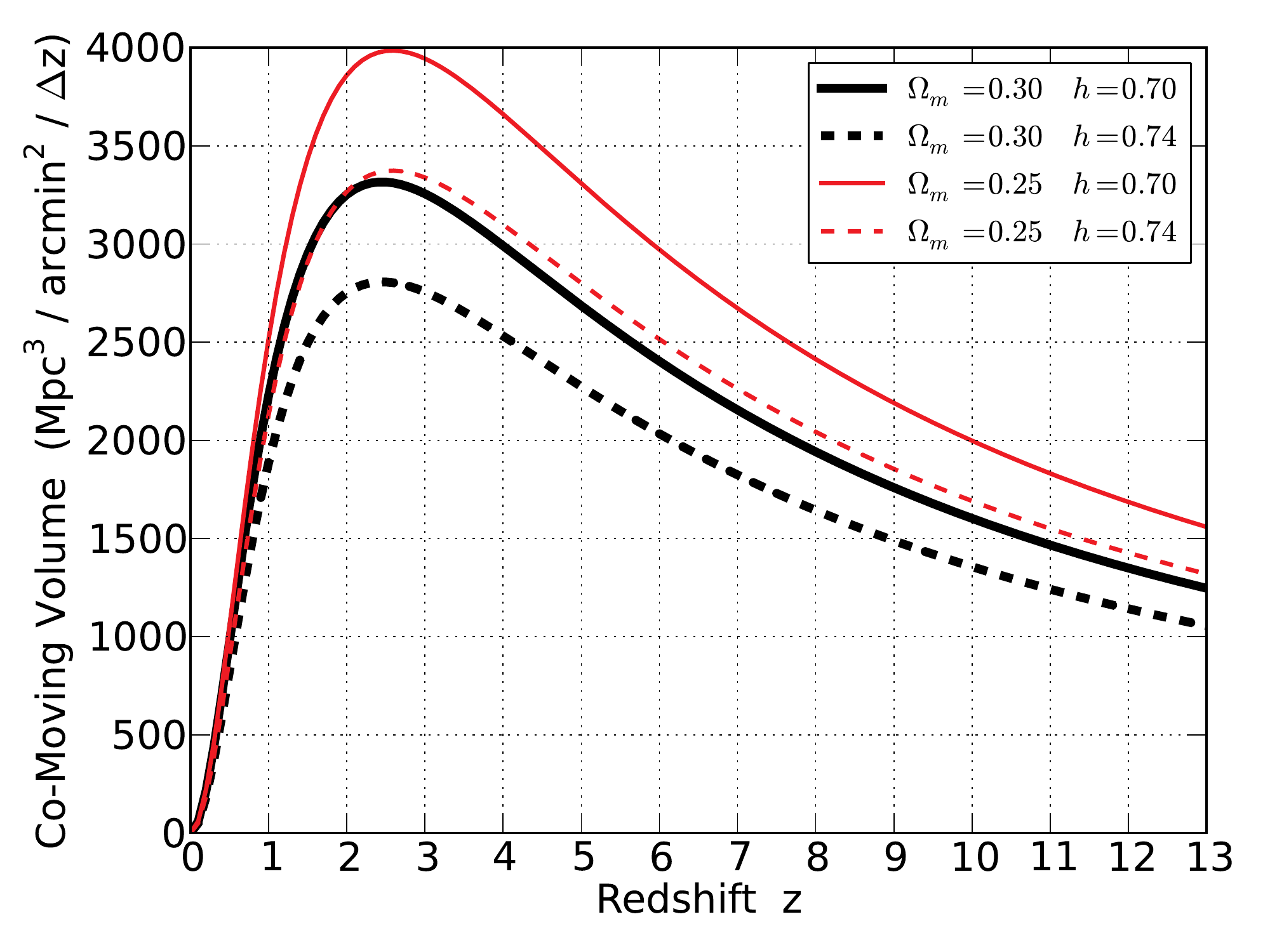}
}
\caption{
\label{fig:volume}
Co-moving volume per unit redshift as a function of redshift
for a range of cosmologies in a flat universe,
provided for reference.
}
\end{figure}

\begin{figure*}
\centerline{
\includegraphics[width = 0.33\textwidth]{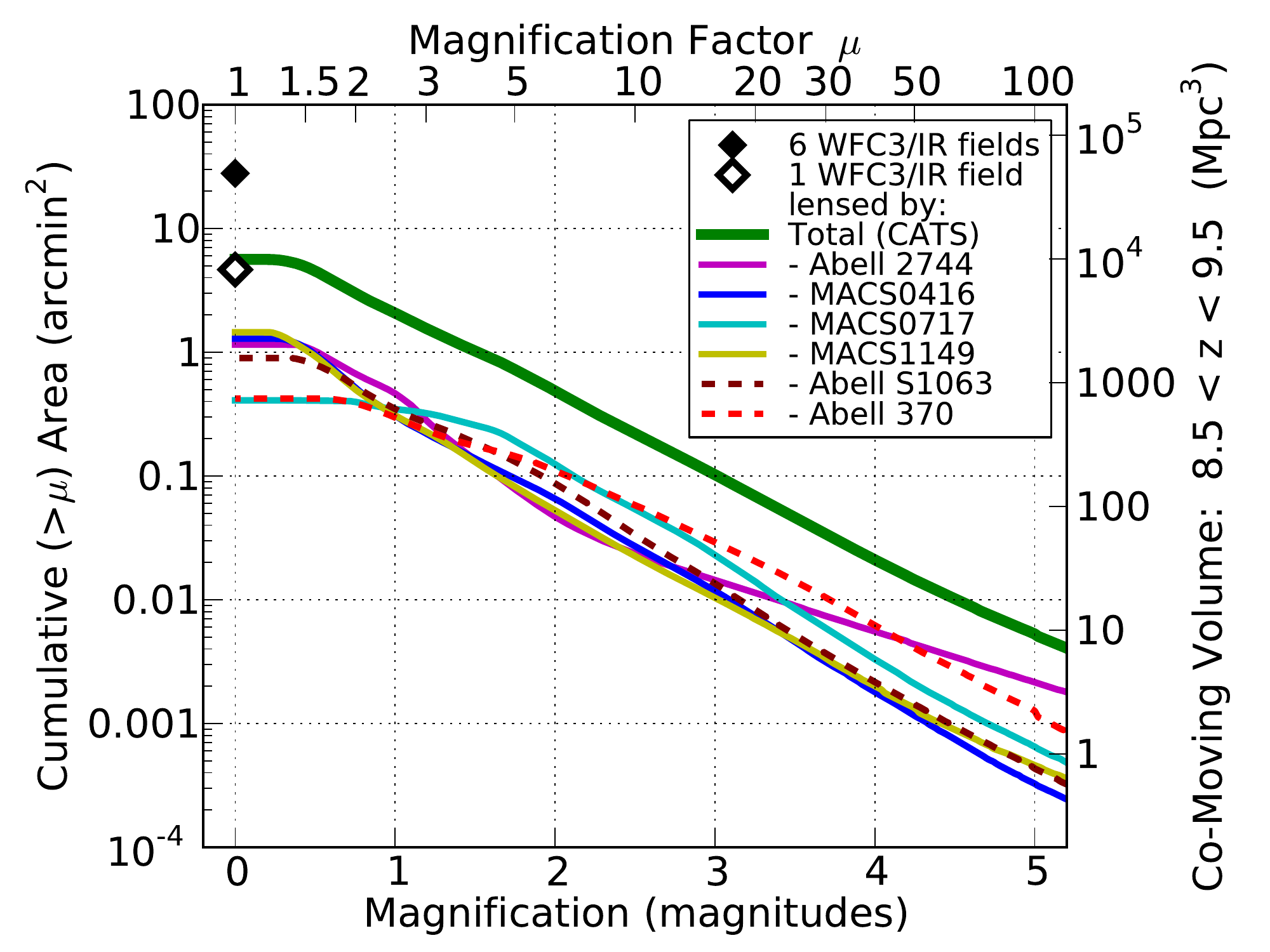}
\includegraphics[width = 0.33\textwidth]{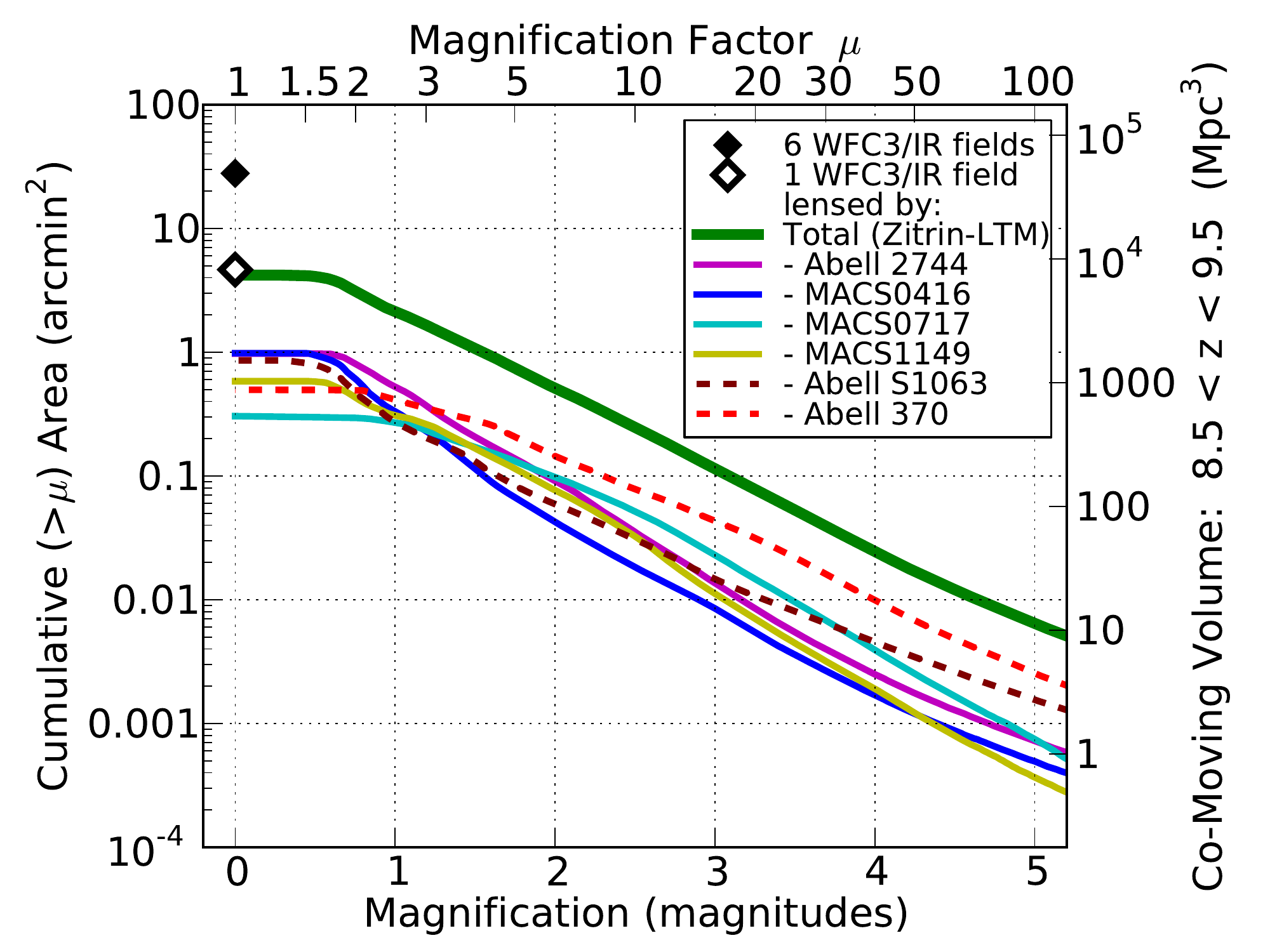}
\includegraphics[width = 0.33\textwidth]{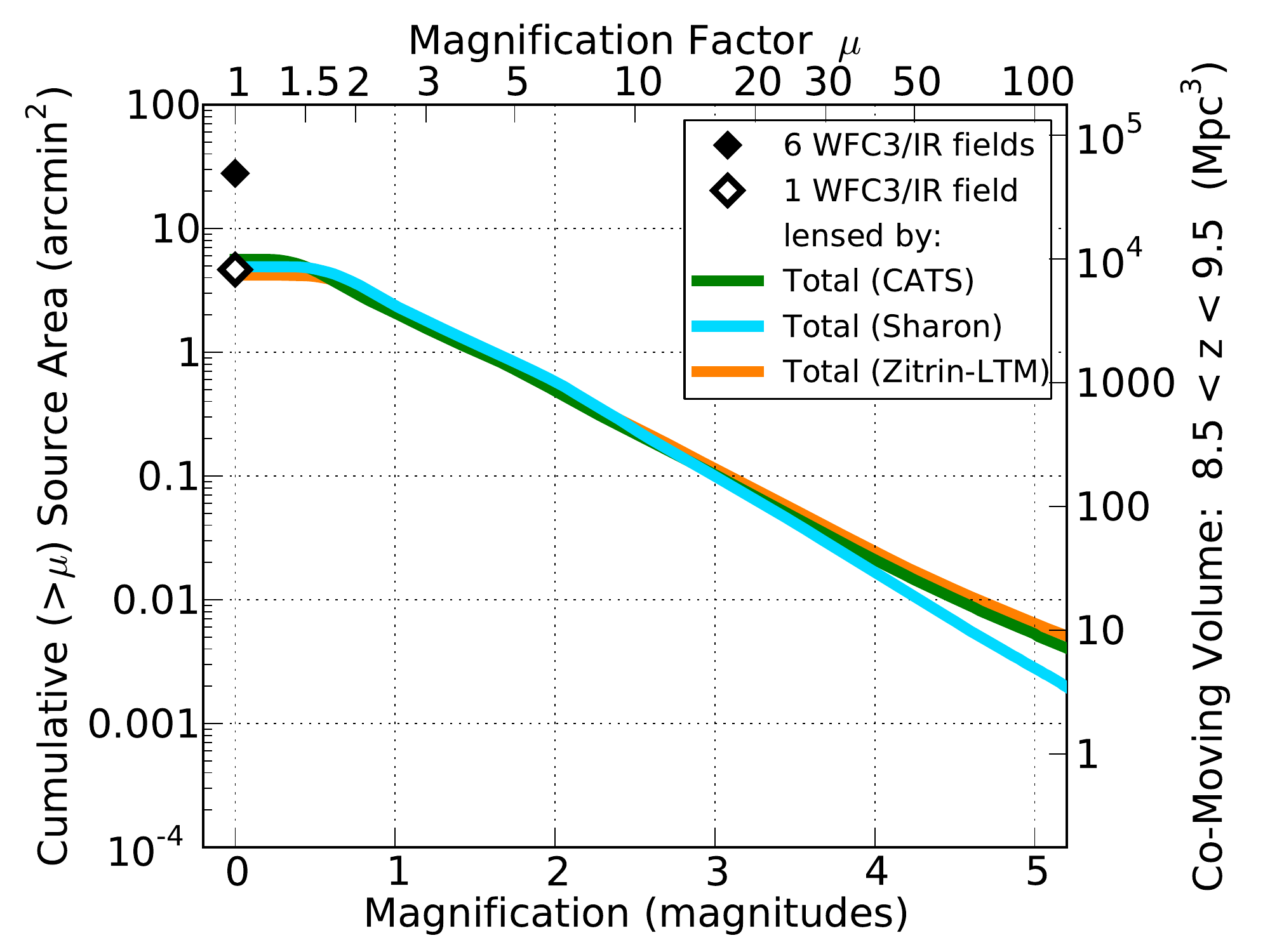}
}
\caption{
\label{fig:volumes}
Cumulative area and corresponding co-moving volume at $z \sim 9$
as a function of magnification for each cluster 
according to the CATS (left) and Zitrin-LTM (center) models.
On both plots we also show both model predictions
for the total volumes lensed by all six clusters.
A corresponding plot for the Sharon models is presented in T.~Johnson et al.~(2014, in preparation).
At right, we compare the total volume predictions for all three models
to highlight their strong agreement.
%
The full survey will yield \~28 arcmin\squared\ (\~50,000 Mpc\cubed\ at $z \sim 9$) in the 6 blank WFC3/IR fields
and \~5 arcmin\squared\ (\~9,000 Mpc\cubed\ at $z \sim 9$) 
of source plane search area in the 6 lensed WFC3/IR fields.
%
These are upper limits as we do not account for area lost due to foreground objects.
The plots also show, for example, that in the lensed fields,
a total of \~1,000 Mpc\cubed\ $z \sim 9$ source plane area
should be magnified by a factor of 6 (\~2 magnitudes) or greater.
}
\end{figure*}

\begin{figure*}
\centerline{
\includegraphics[width = 0.33\textwidth]{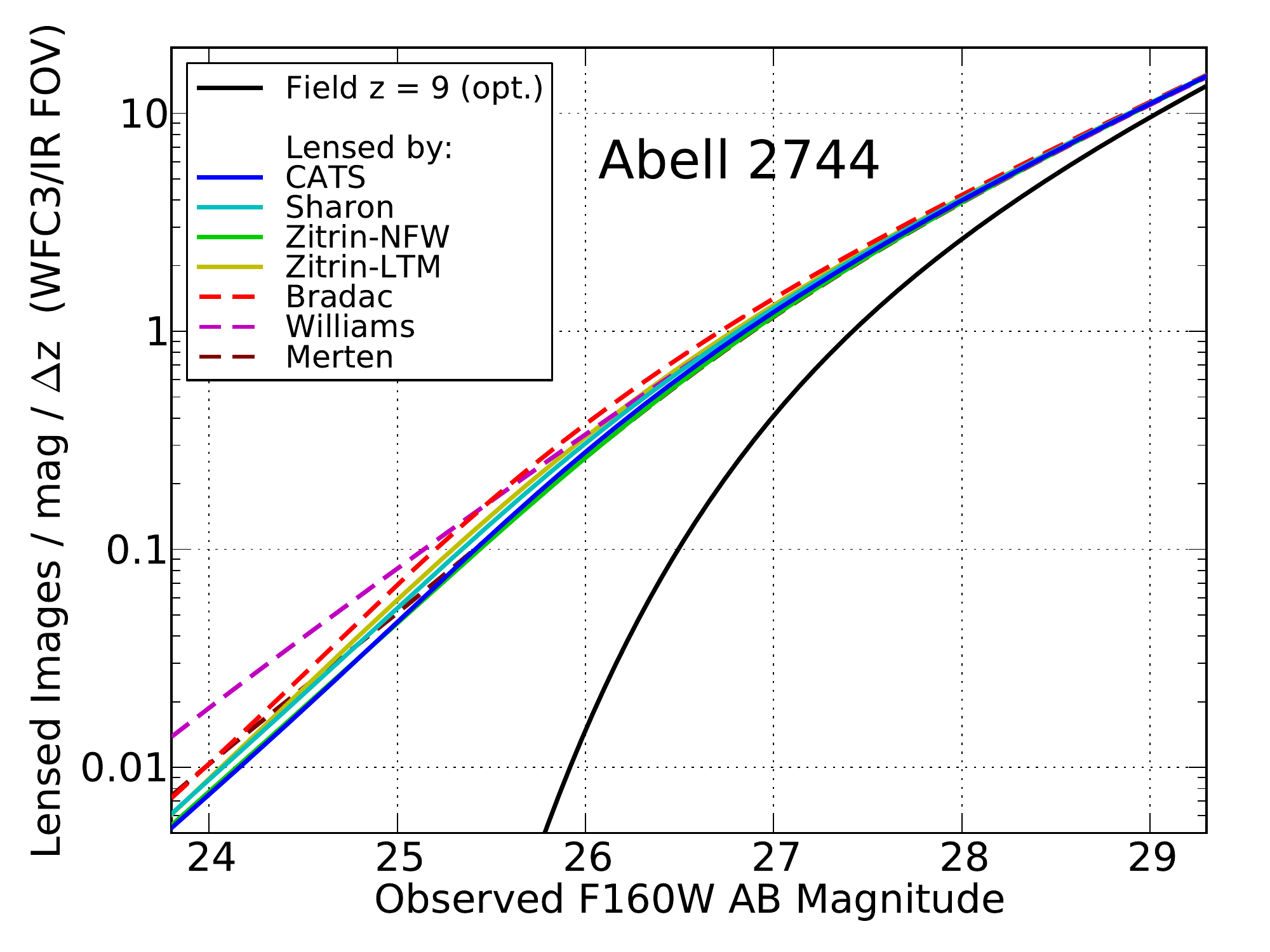}
\includegraphics[width = 0.33\textwidth]{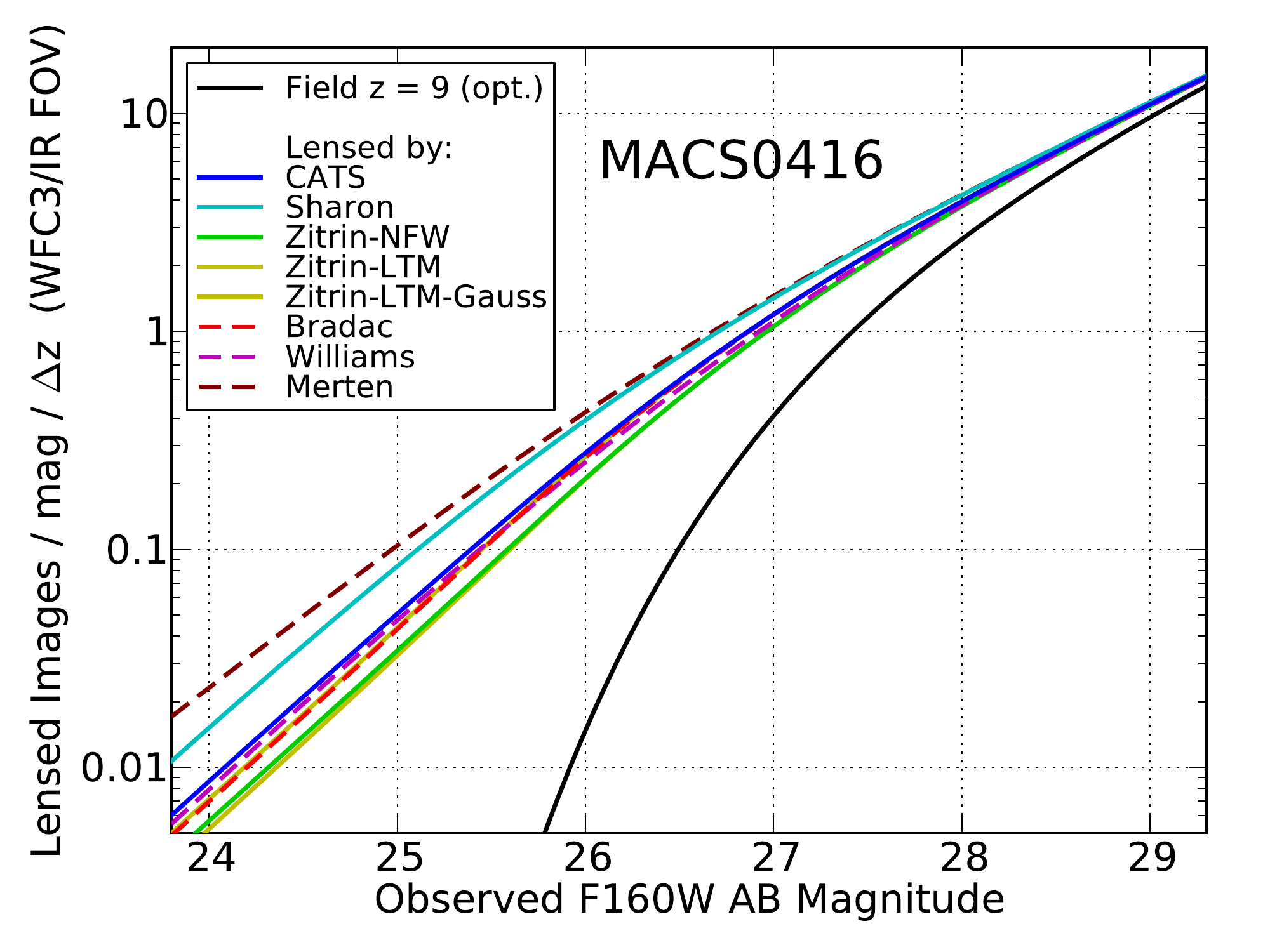}
\includegraphics[width = 0.33\textwidth]{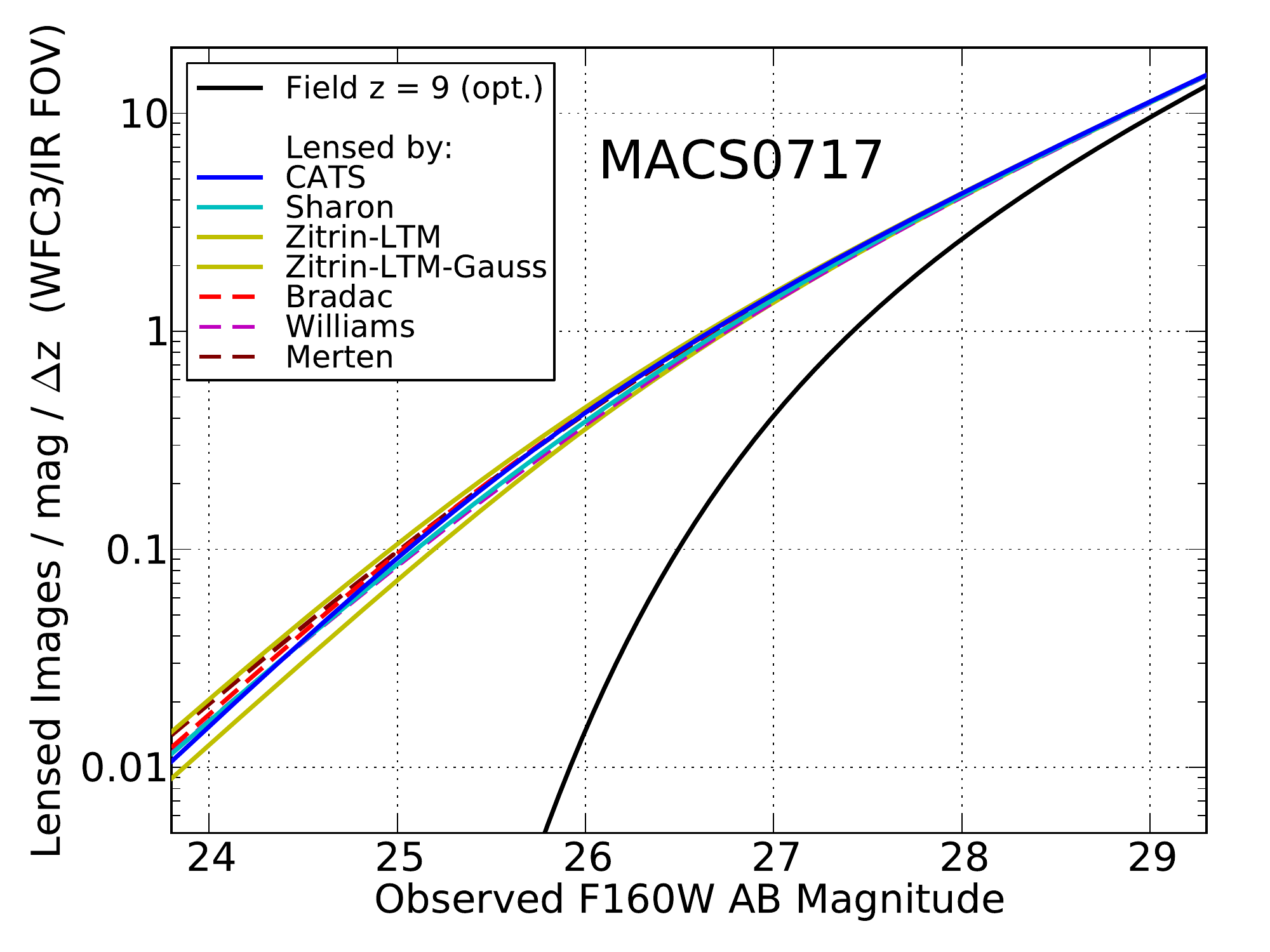}
}
\centerline{
\includegraphics[width = 0.33\textwidth]{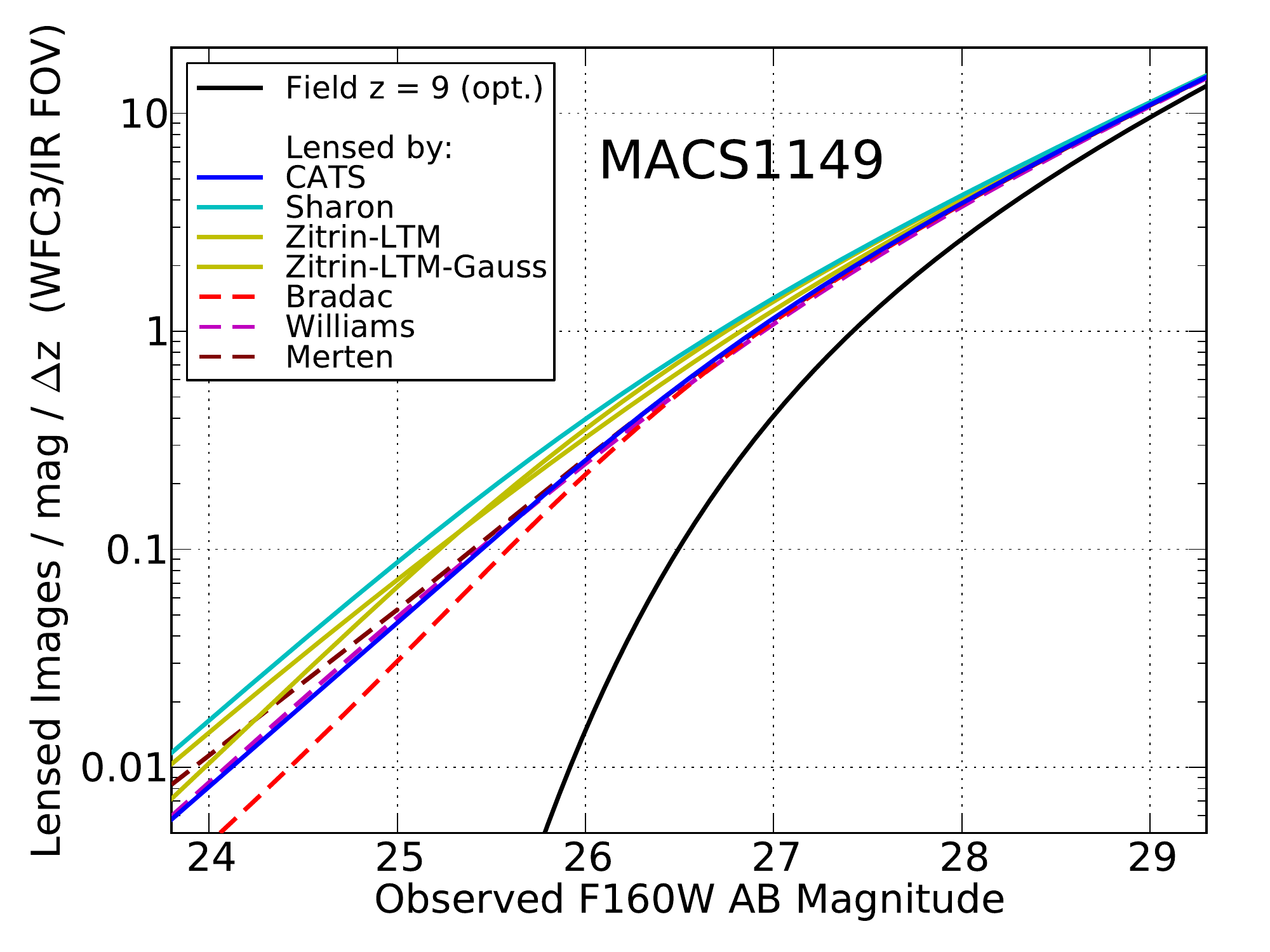}
\includegraphics[width = 0.33\textwidth]{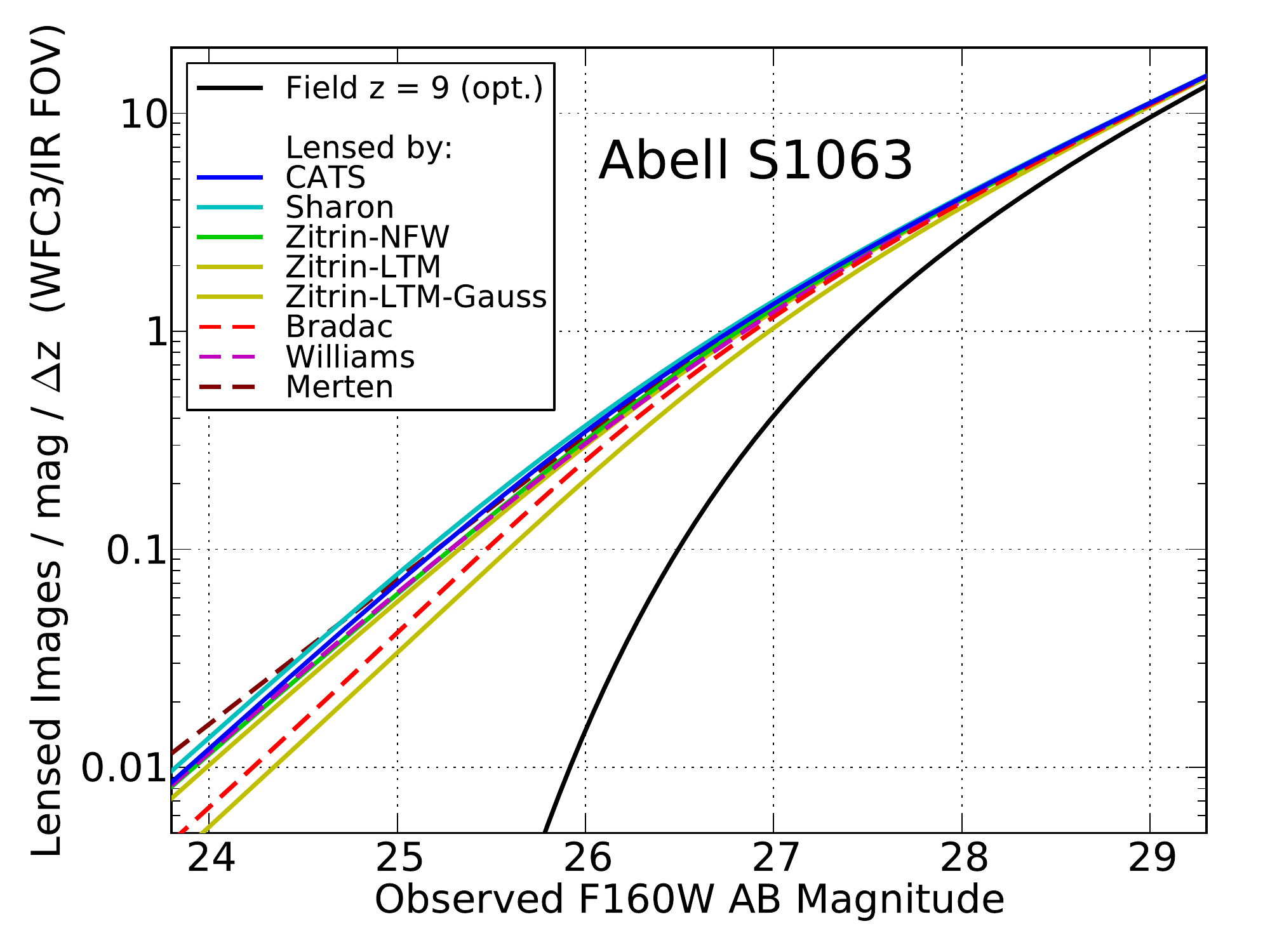}
\includegraphics[width = 0.33\textwidth]{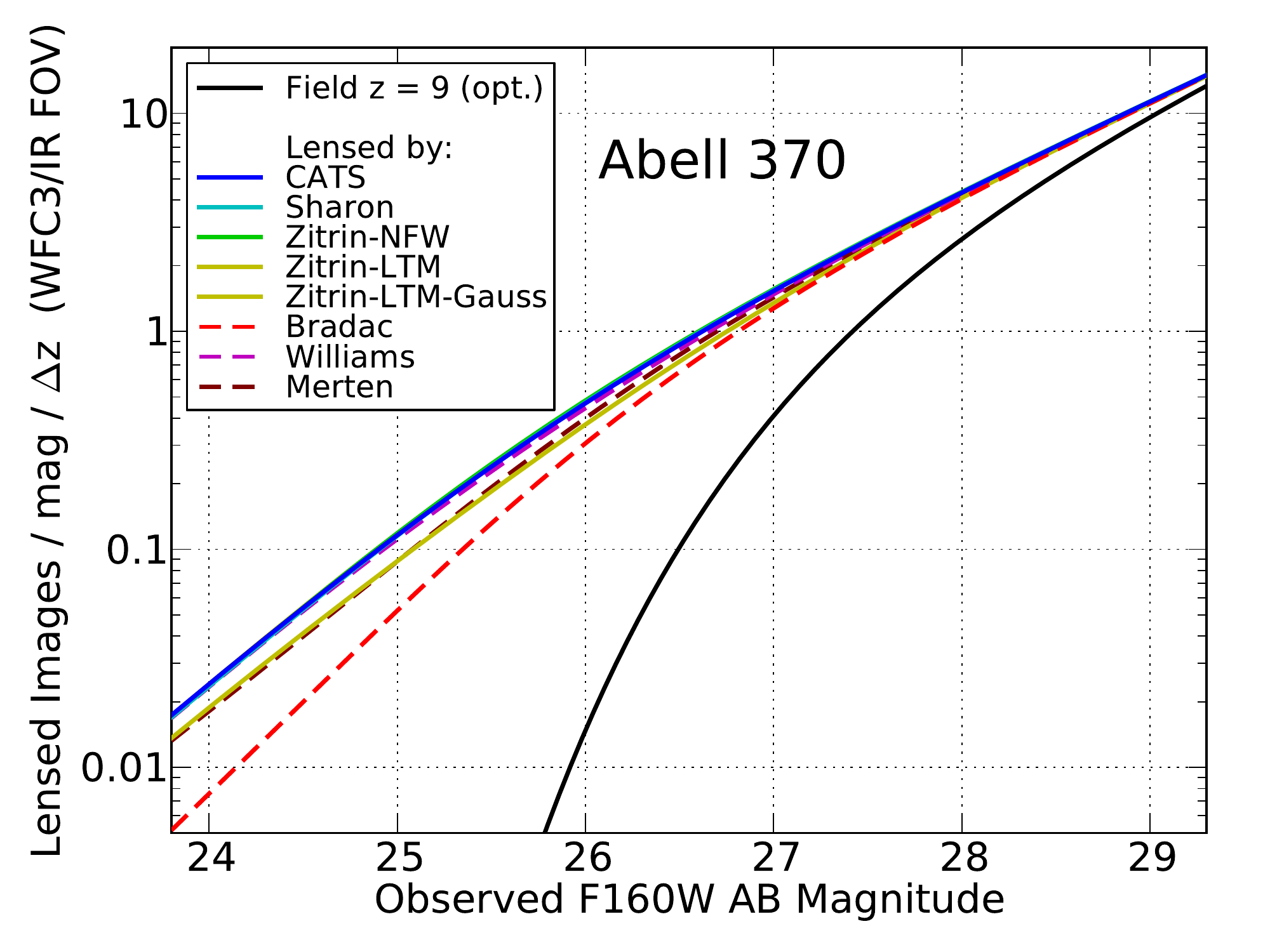}
}
\caption{
\label{fig:lensLFs}
Optimistic numbers of lensed images of $z \sim 9$ galaxies expected for each Frontier Field.
We adopt an optimistic evolution of the \cite{Bradley12b} LF from $z = 8$ to 9 (\S\ref{sec:fieldLF})
and lens this through every lens model submitted,
yielding numbers of lensed images 
per unit magnitude and unit redshift
within the WFC3/IR FOV (4.65 arcmin\squared) as a function of observed magnitude.
Incompleteness and contamination are not considered here; 
these must be estimated and accounted for in any search for high-redshift galaxies.
}
\end{figure*}

\begin{figure*}
\centerline{
\includegraphics[width = 0.33\textwidth]{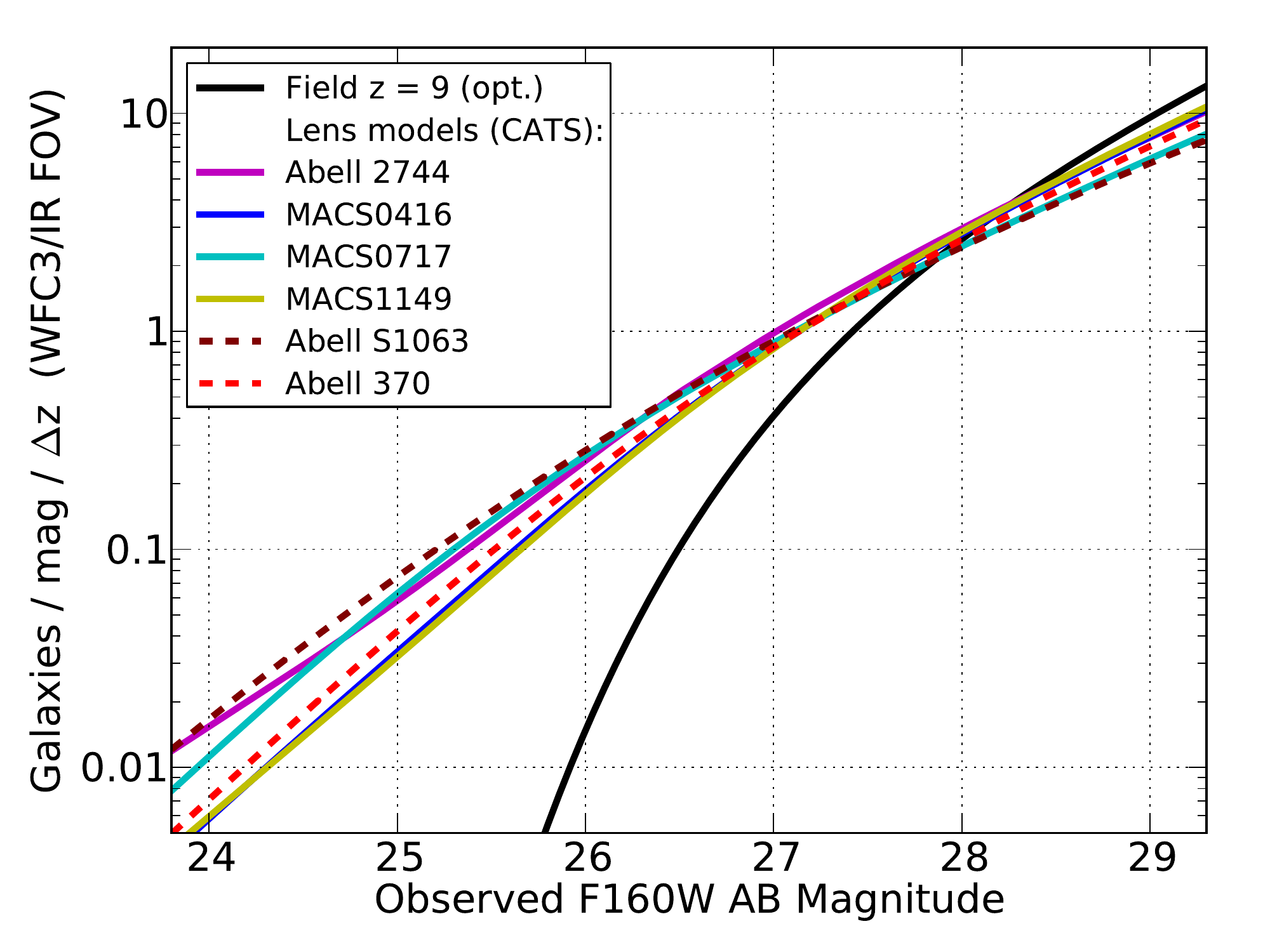}
\includegraphics[width = 0.33\textwidth]{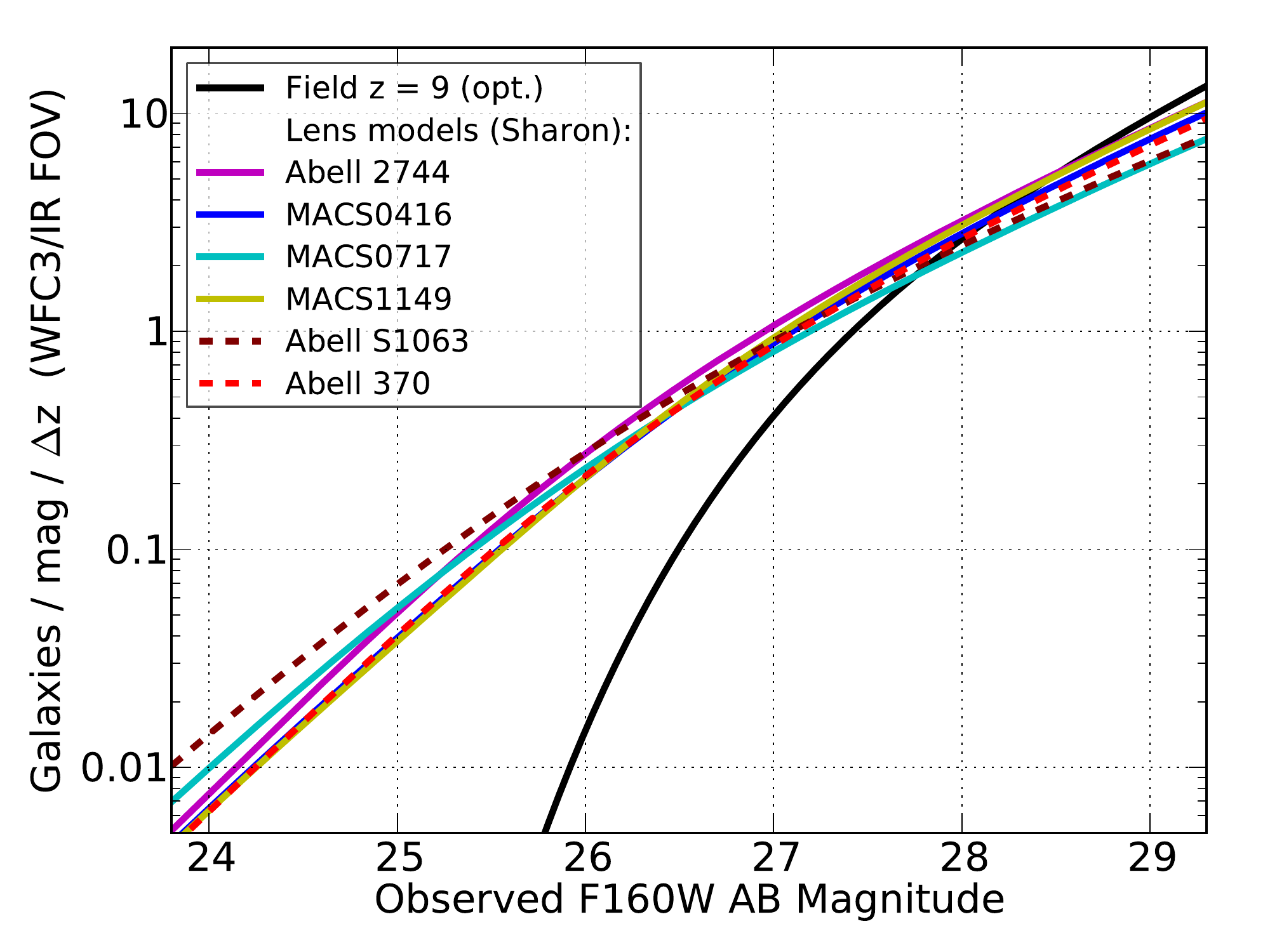}
\includegraphics[width = 0.33\textwidth]{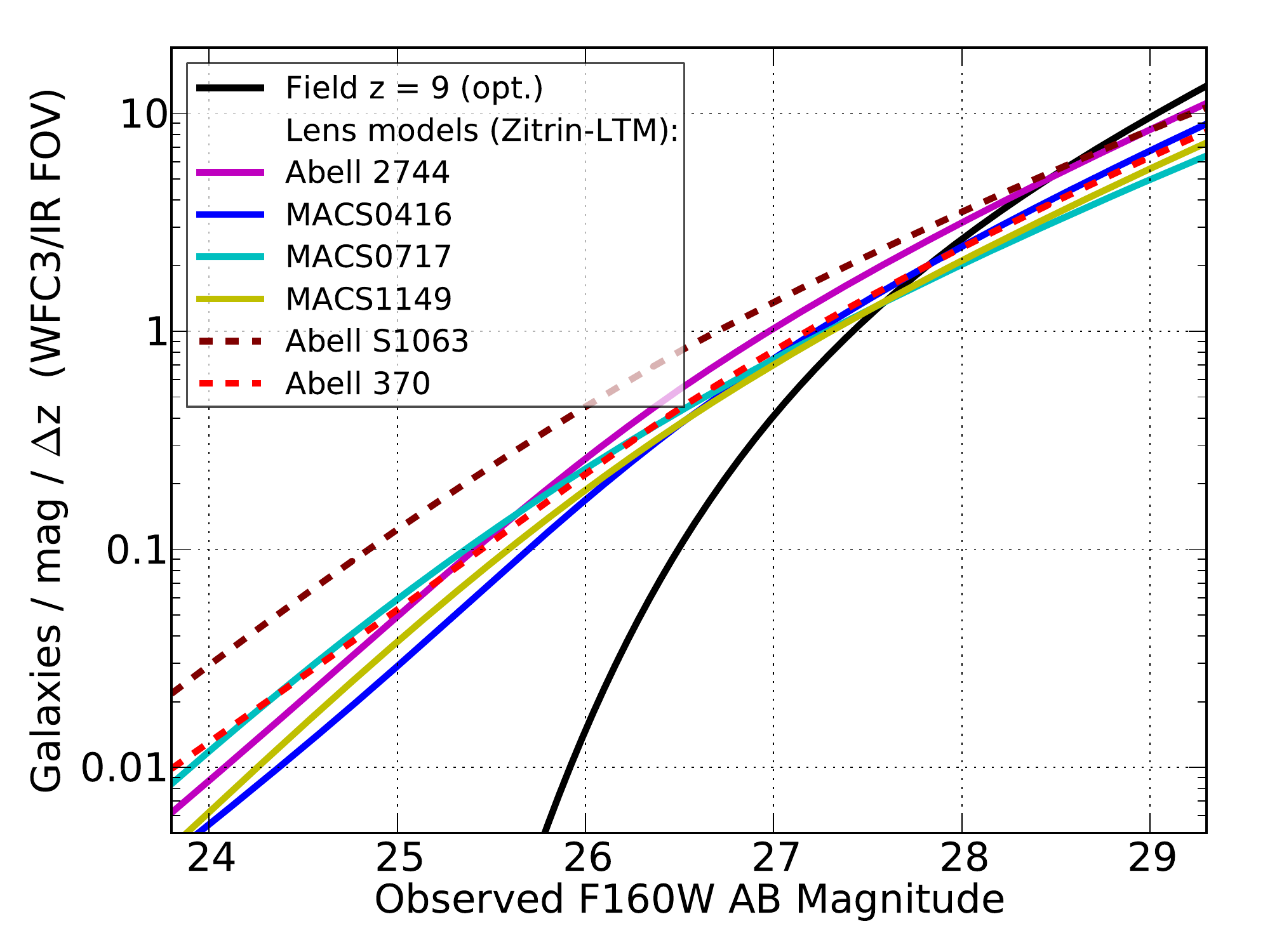}
}
\caption{
\label{fig:lensLFcats}
Optimistic numbers of $z \sim 9$ galaxies expected to be lensed by each Frontier Fields cluster 
according to the CATS, Sharon, and Zitrin-LTM models.
In these plots, each galaxy lensed to form multiple images is only counted once as a single galaxy:
we consider only the most strongly magnified galaxy.
(For comparison, Figure \ref{fig:lensLFs} tallied the total numbers of lensed images.)
}
\end{figure*}

\begin{figure*}
\centerline{
\includegraphics[width = 0.49\textwidth]{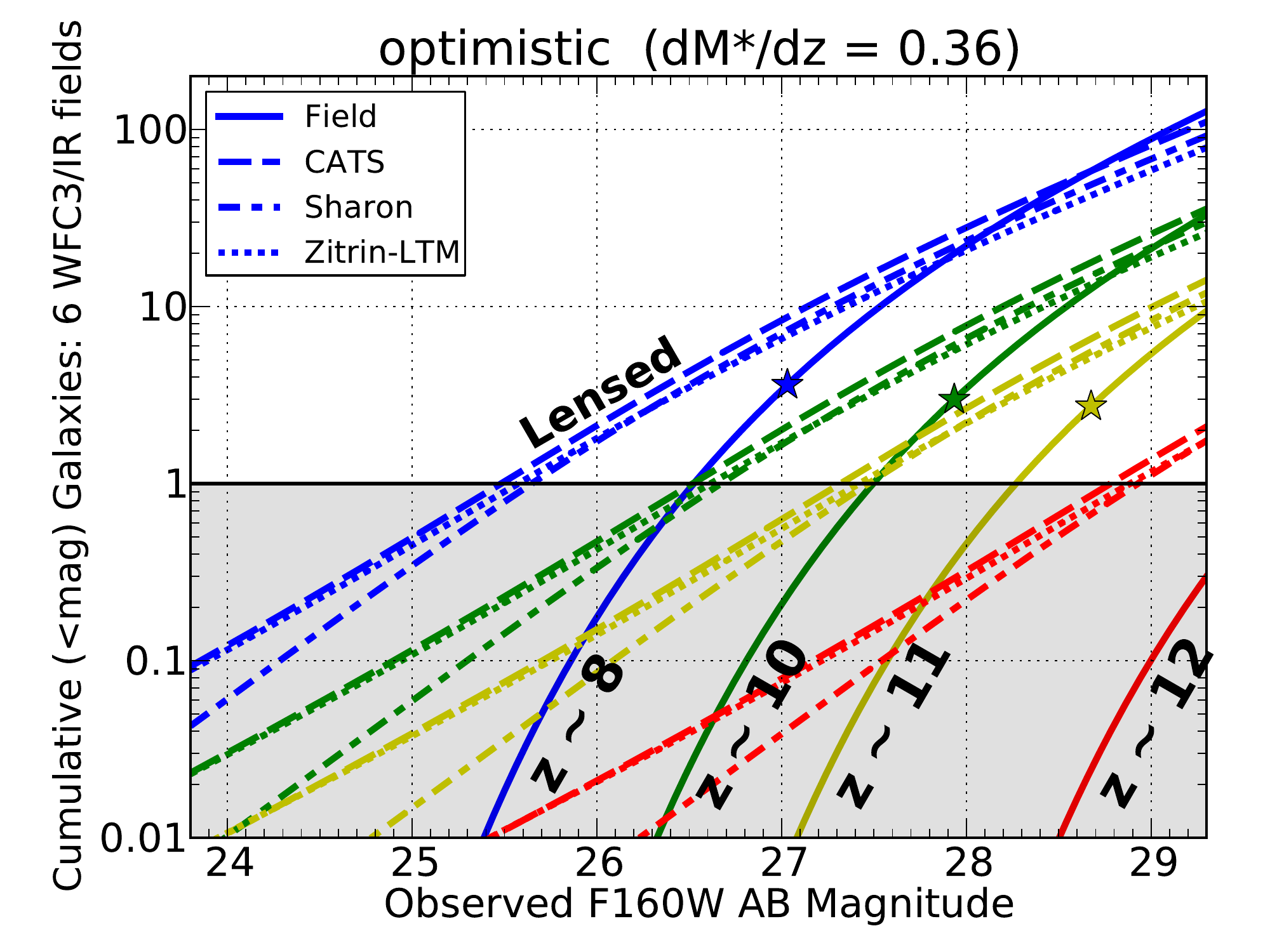}
\includegraphics[width = 0.49\textwidth]{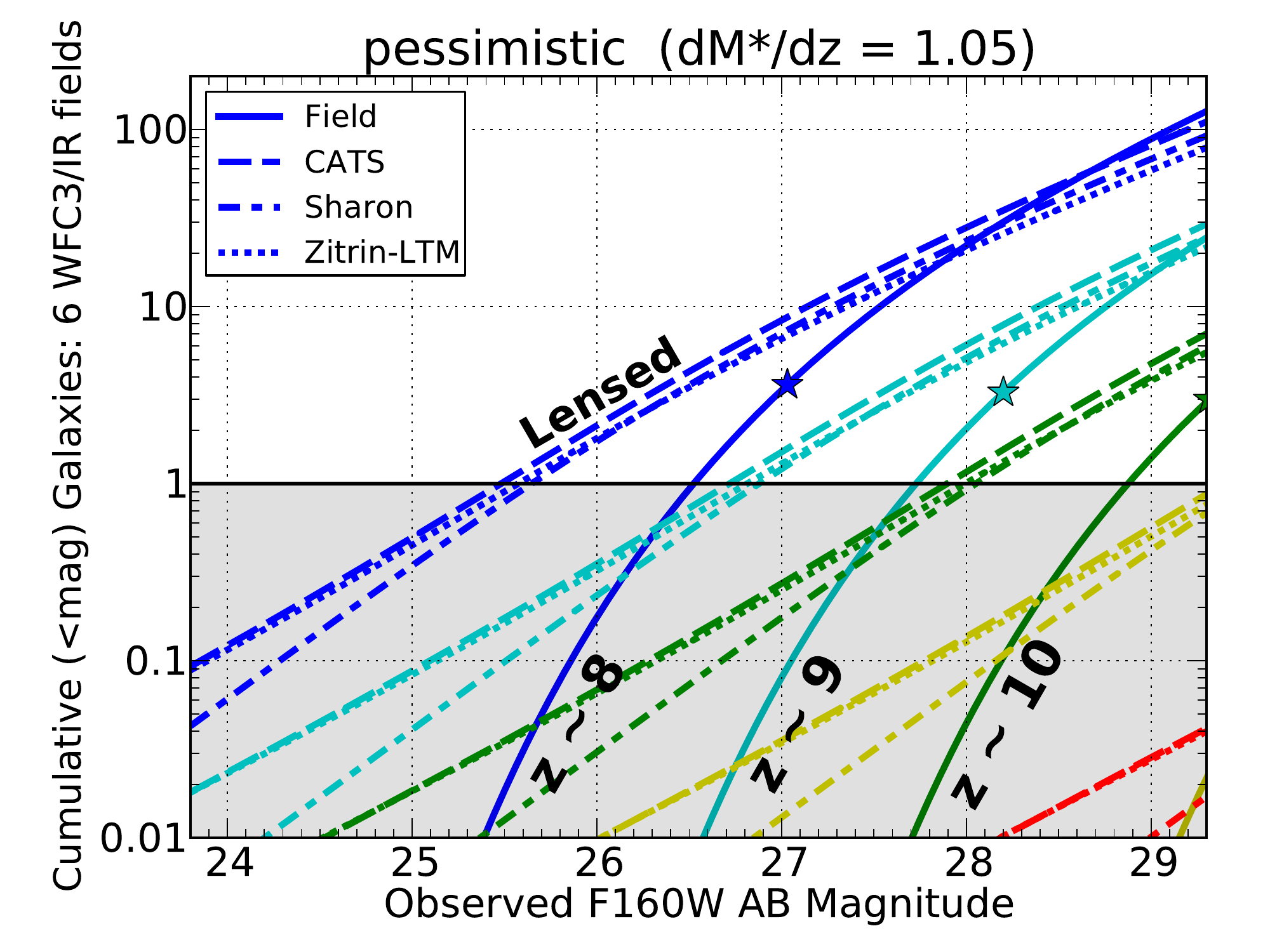}
}
\caption{
\label{fig:FF6}
Total number counts of high-redshift ($z \sim 8$ -- 12) galaxies
as a function of magnitude limit
expected from all six 
blank fields (solid lines)
and cluster lensed fields (dashed lines)
according to the CATS, Sharon, and Zitrin-LTM models.
On each solid line, we mark our assumed $M^*(z)$ with a star.
We show both optimistic (left) and pessimistic (right) expectations.
}
\end{figure*}

\begin{figure*}
\centerline{
\includegraphics[width = 0.49\textwidth]{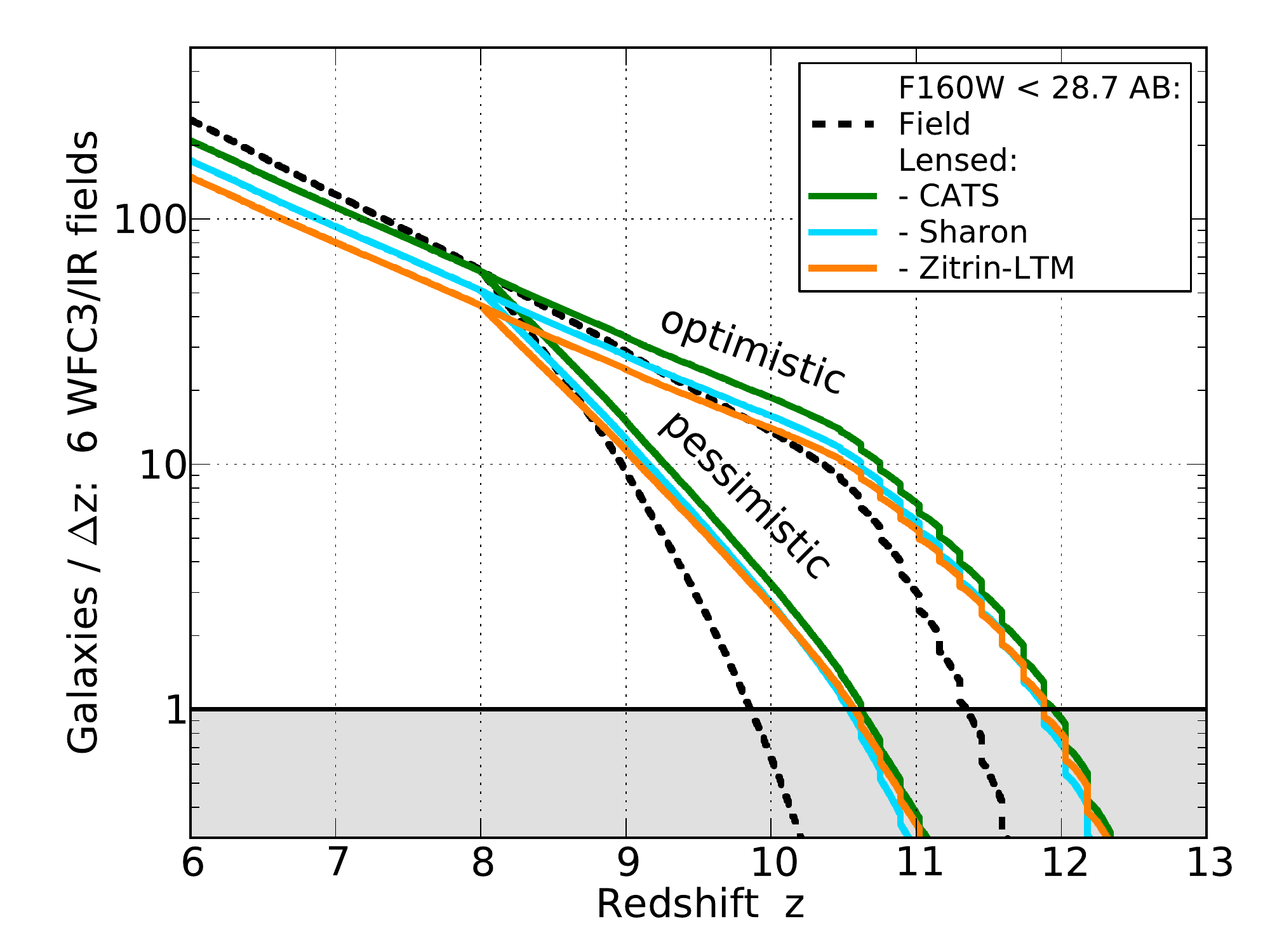}
\includegraphics[width = 0.49\textwidth]{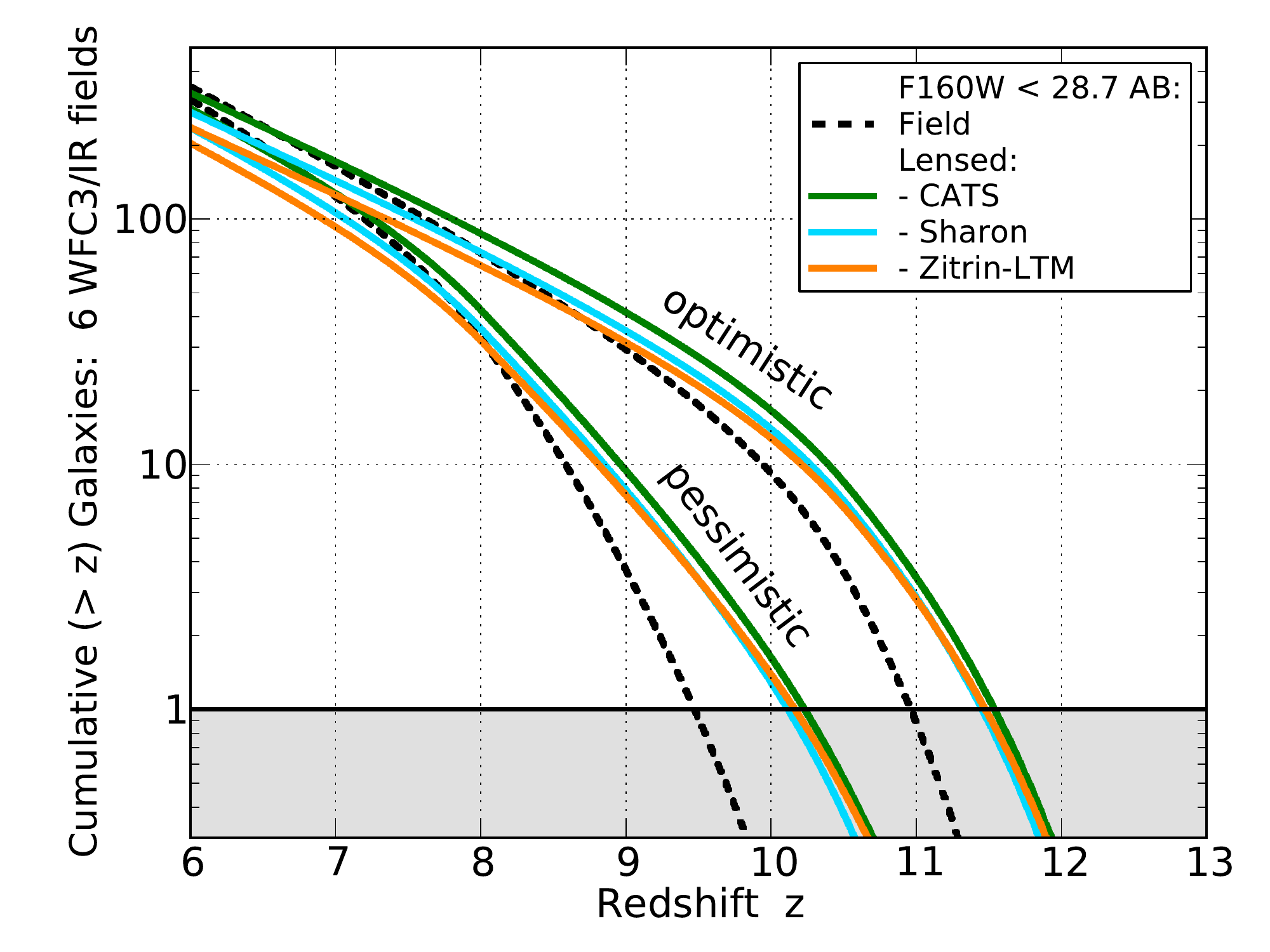}
}
\caption{
\label{fig:FF6cum}
Total high-redshift number counts expected brighter than the F160W magnitude limit 28.7 AB
for all six blank and lensed fields
according to the CATS, Sharon, and Zitrin-LTM lens models.
Left: total counts in each redshift bin of width $\Delta z = 0.1$.
Right: total cumulative counts as a function of redshift.
For example, we expect as many as \~40 lensed $z > 9$ galaxies.
We show both optimistic and pessimistic $z > 8$ evolution
as described in the text and Figure \ref{fig:FF6}.
}
\end{figure*}

\begin{deluxetable}{lrrrrrrrr}
\tablecaption{\label{tab:counts}Approximate numbers of high-redshift galaxies expected in the full survey
(six ``blank'' fields and six lensed)
if complete to F160W $<$ 28.7 AB.}
\tablewidth{\columnwidth}
\tablehead{
\colhead{}&
\colhead{}&
\multicolumn{3}{c}{Optimistic\supa}&
\colhead{}&
\multicolumn{3}{c}{Pessimistic\supa}\\
\colhead{Redshift\supb}&
\colhead{}&
\colhead{Field}&
\colhead{Lensed}&
\colhead{Total}&
\colhead{}&
\colhead{Field}&
\colhead{Lensed}&
\colhead{Total}
}
\startdata
~~$z \sim 8$\supc && 62 & 61 & 123 && 62 & 61 & 123\\
~~$z \sim 9$ && 29 & 33 & 52 && 9 & 15 & 24\\
~~$z \sim 10$ && 14 & 19 & 33 && 0.6\hspace{-0.09in} & 3 & 4\\
~~$z \sim 11$ && 3 & 7 & 10 && 0 & 0.4\hspace{-0.09in} & 0\\
~~$z \sim 12$ && 0 & 1 & 1 && 0 & 0 & 0\\
~~$z > 9$ && 29 & 42 & 71 && 3.5\hspace{-0.09in} & 9 & 13
\enddata
\tablenotetext{1}{``Optimistic'' and ``Pessimistic'' refer to $z > 8$ LF extrapolations.}
\tablenotetext{2}{As elsewhere in this manuscript, $z \sim 8$ corresponds to $7.5 < z < 8.5$, etc.}
\tablenotetext{3}{Uncertainties in $z \sim 8$ counts are \~30\%.}
\end{deluxetable}

\begin{figure*}
\centerline{
\includegraphics[width = 0.49\textwidth]{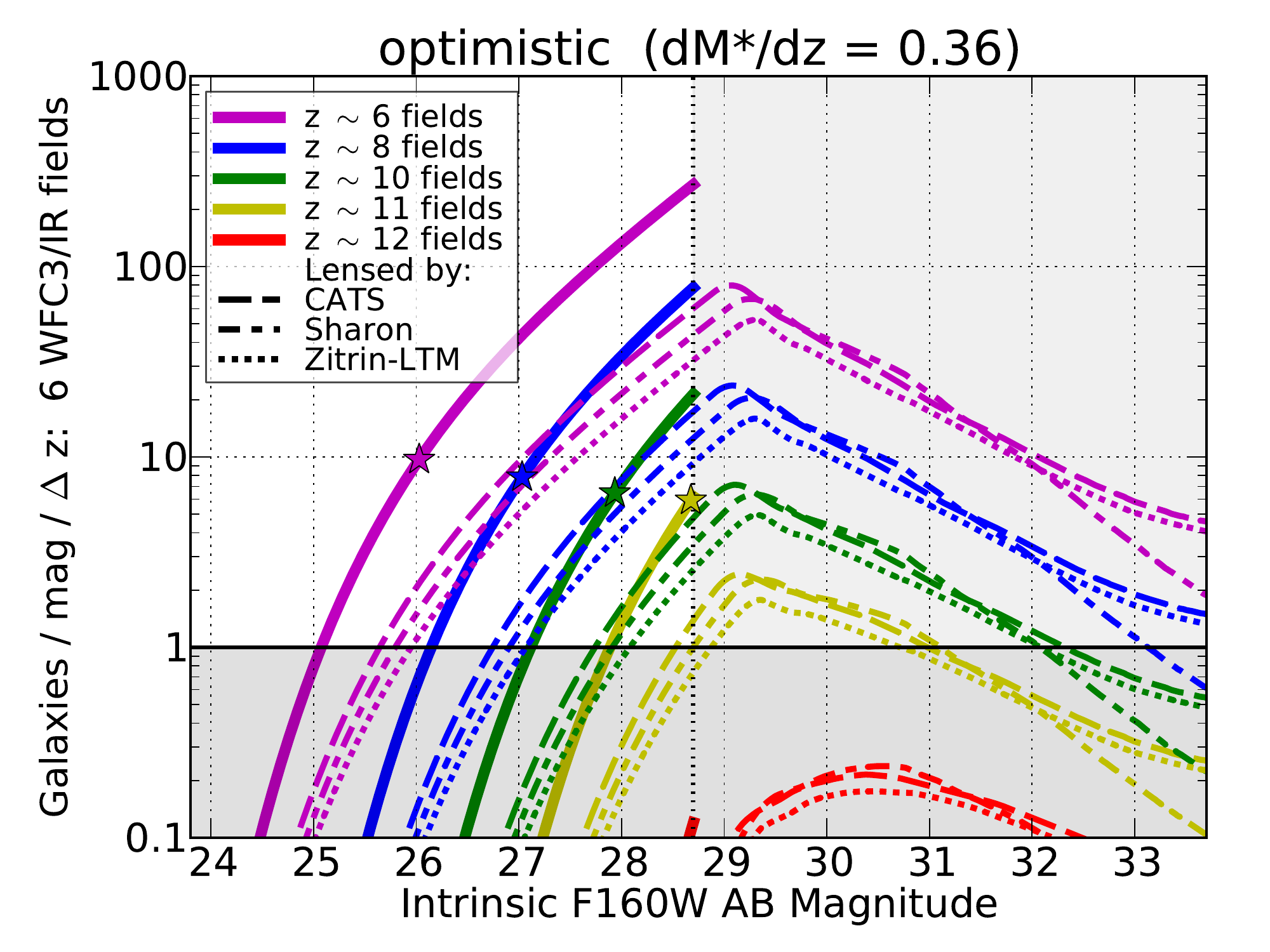}
\includegraphics[width = 0.49\textwidth]{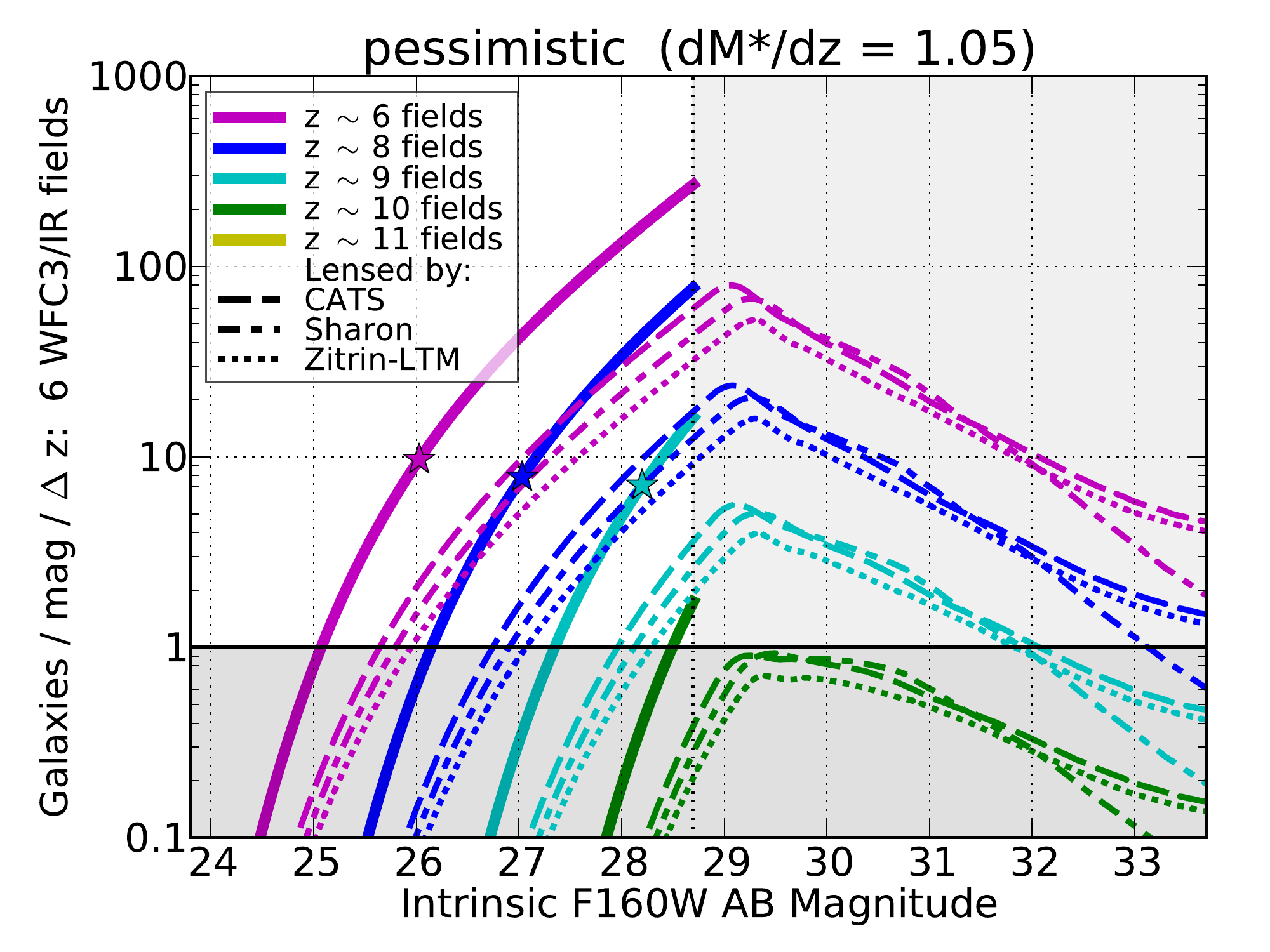}
}
\caption{
\label{fig:FF6int}
Expected numbers of galaxies observed with F160W $<$ 28.7 AB
plotted as a function of intrinsic (unlensed) magnitude.
We mark assumed $M^*(z)$ values with stars.
Lensed counts peak at a slightly fainter magnitude (29.0)
since magnifications are typically \~0.3 mag and greater across the entire WFC3/IR field
(according to the CATS, Sharon, and Zitrin-LTM models).
Number counts decline for still fainter intrinsic magnitudes because higher magnifications are more rare.
Significant numbers of lensed galaxies are expected in the Frontier Fields
to magnitudes fainter than the UDF12 5\sig\ point source detection limit 
F160W $<$ 29.45 AB \citep{Koekemoer13}.
%
}
\end{figure*}

\begin{figure*}
\centerline{
\includegraphics[width = 0.40\textwidth]{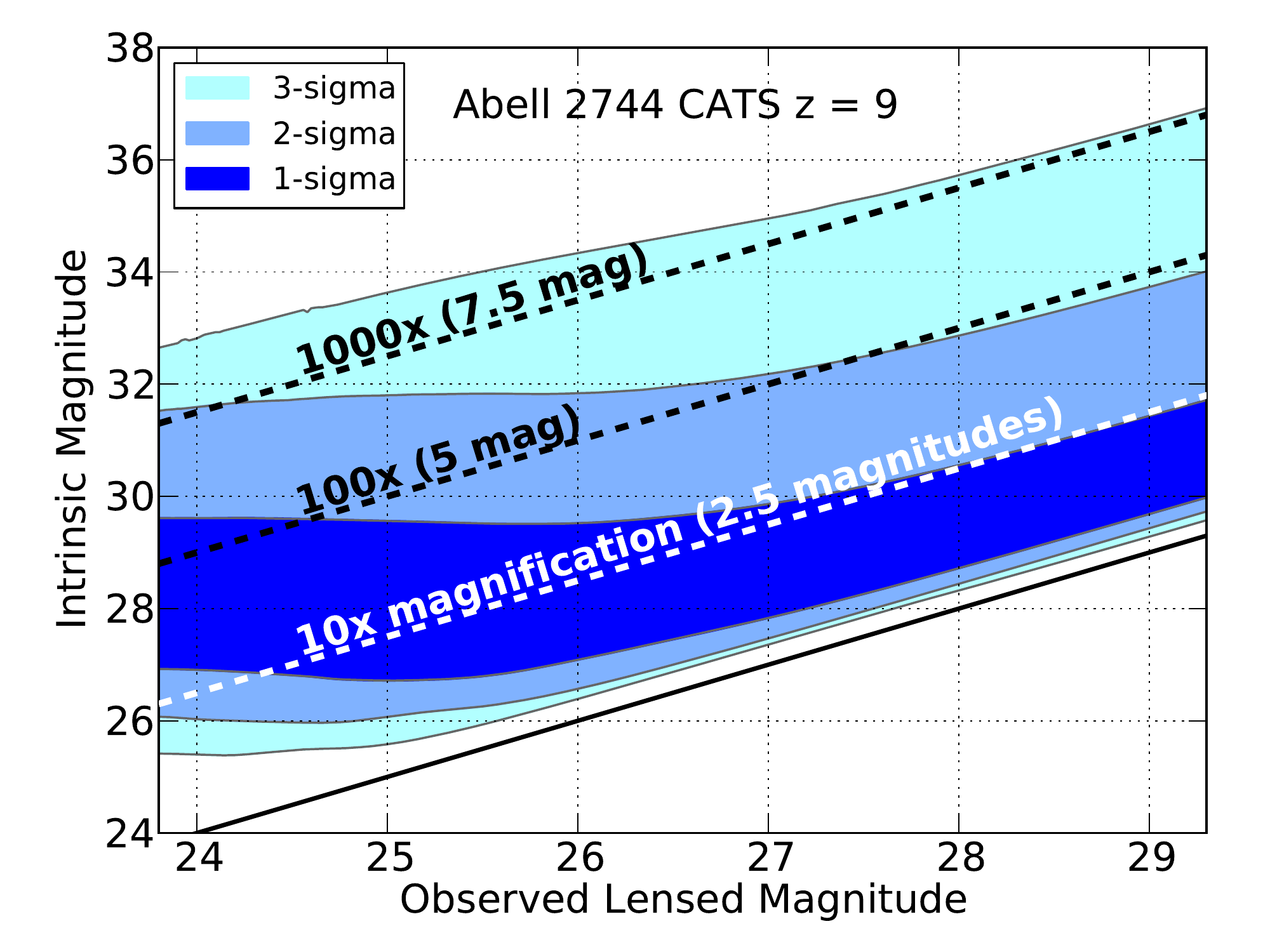}
\includegraphics[width = 0.40\textwidth]{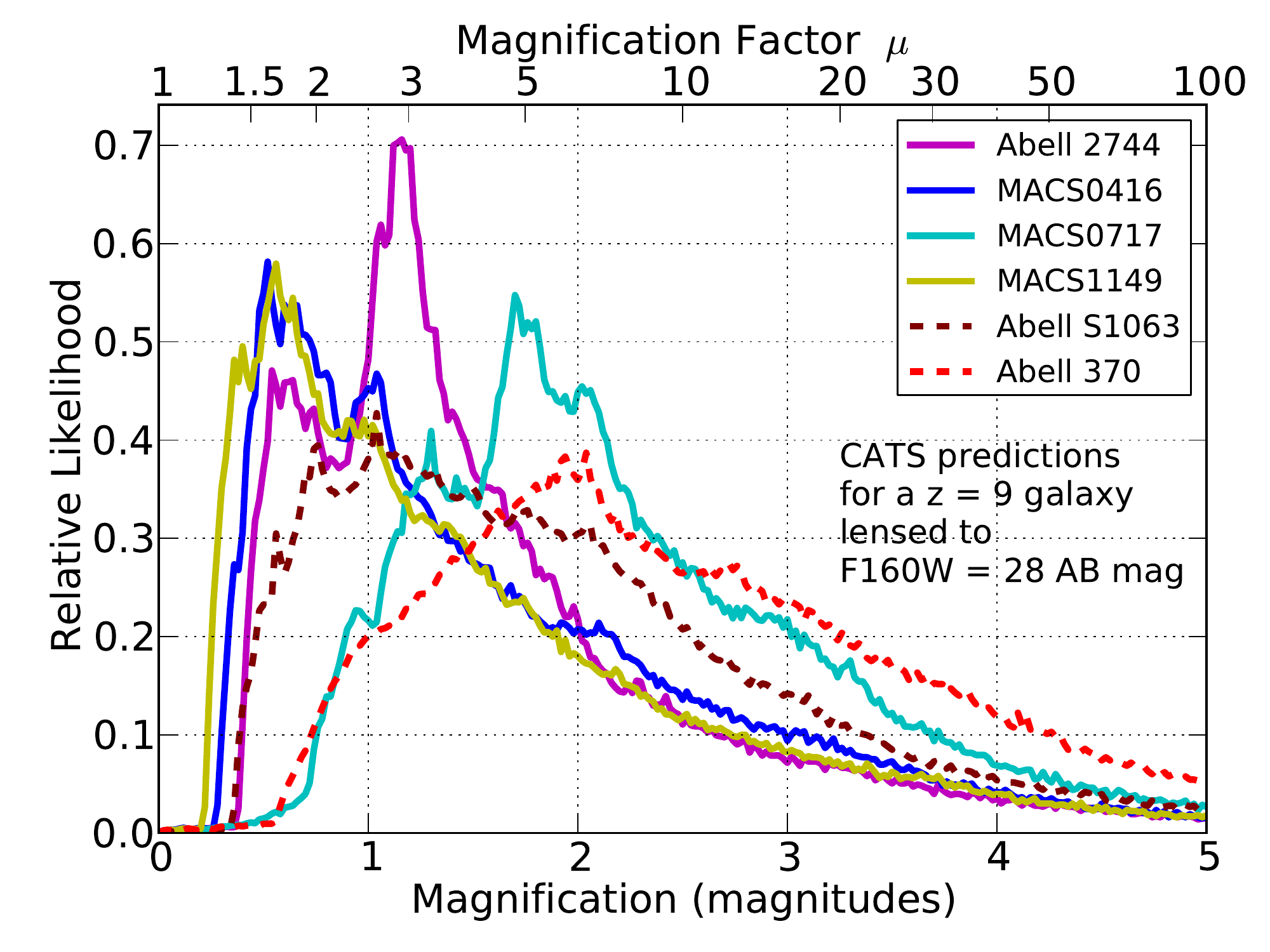}
}
\centerline{
\includegraphics[width = 0.40\textwidth]{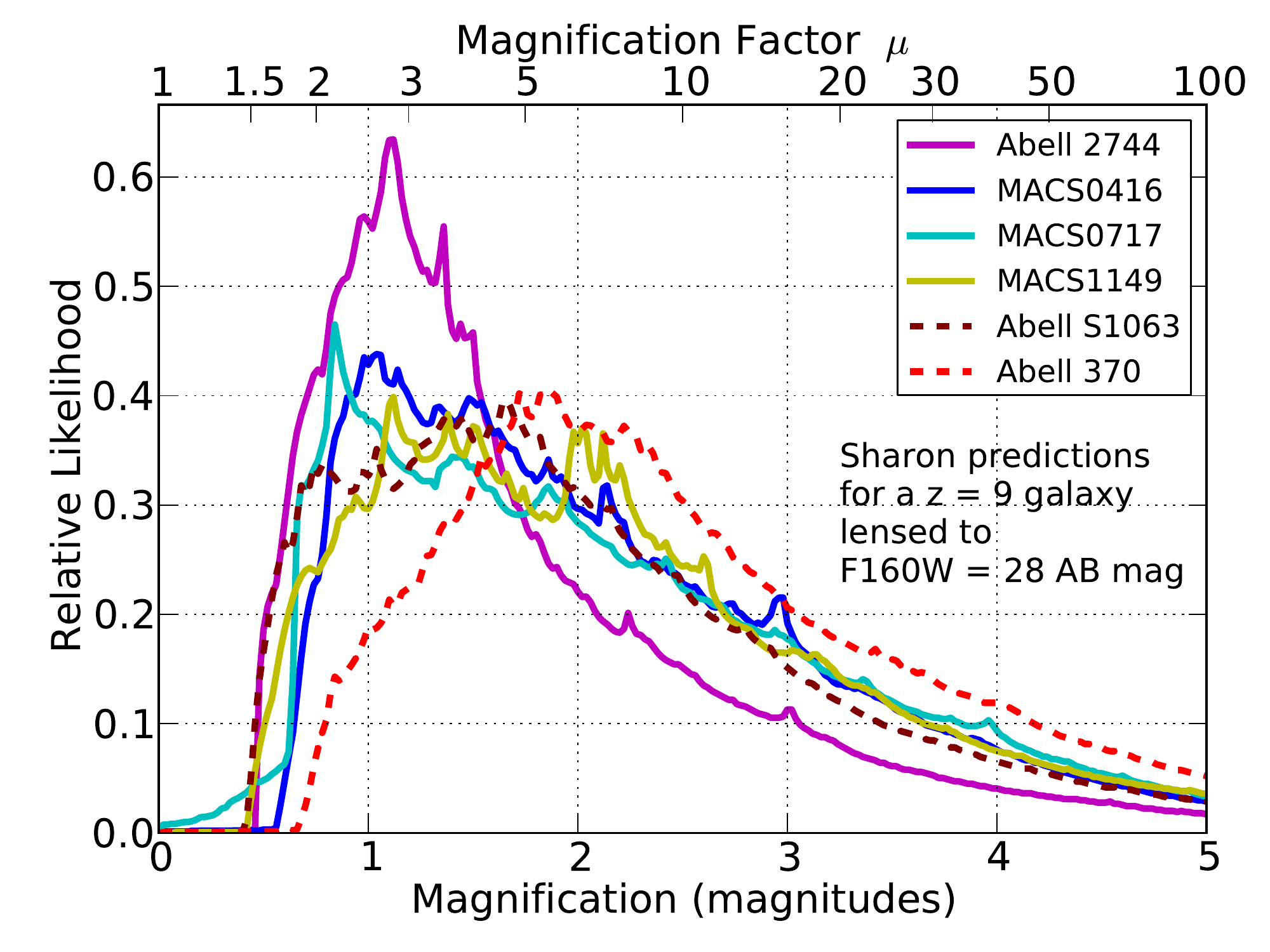}
\includegraphics[width = 0.40\textwidth]{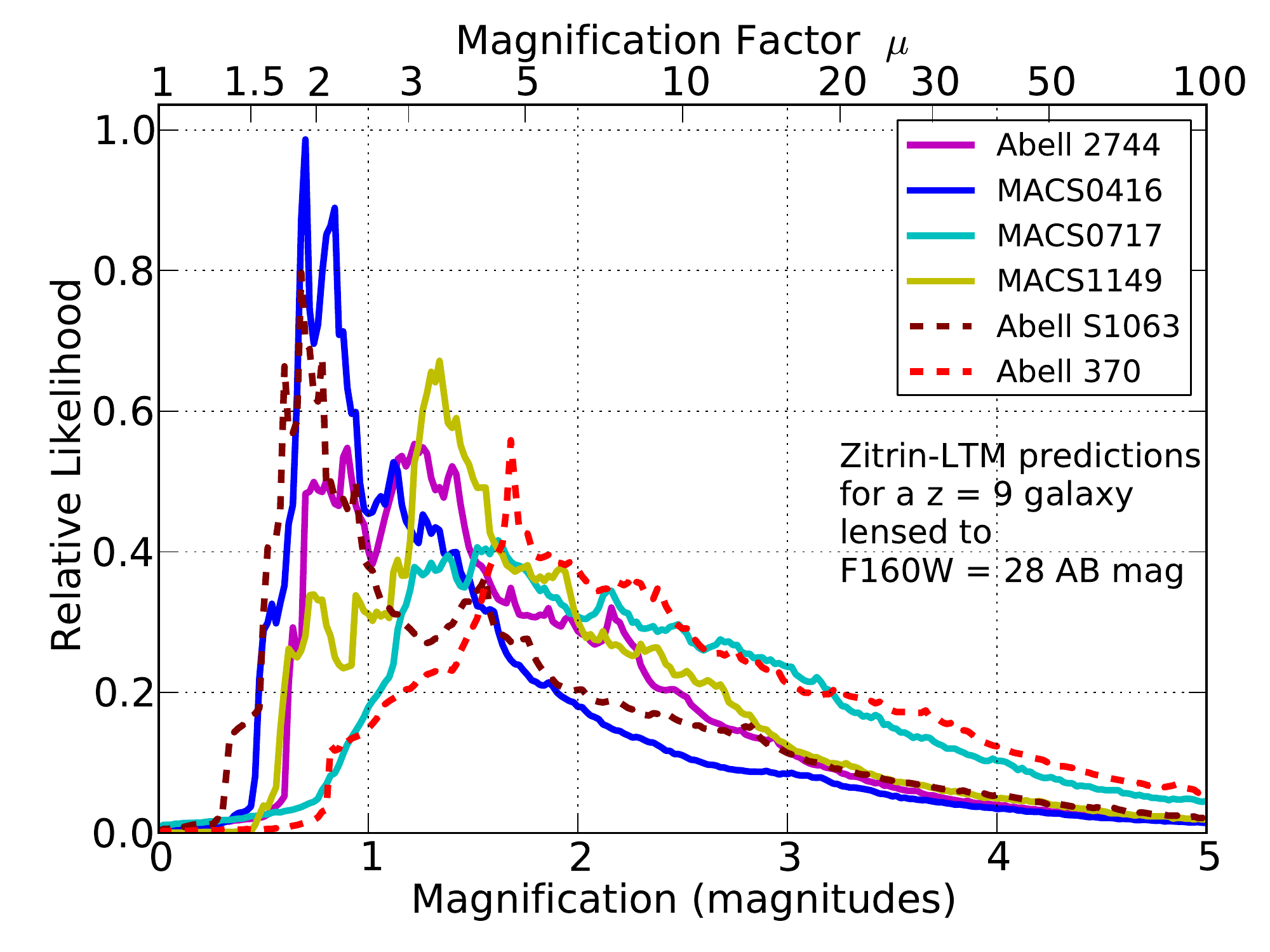}
}
\caption{
\label{fig:mobsint}
Top left: Magnification likelihoods as a function of observed lensed magnitude for $z = 9$ galaxies  
lensed by Abell 2744 according to the CATS model and assuming ``optimistic'' $z > 8$ evolution.
At lower redshifts, galaxies require higher magnifications less often to be observed at brighter magnitudes.
The corresponding plots for other clusters are qualitatively similar with vertical shifts.
Top right: Magnification likelihood distributions at a fixed observed magnitude 
(F160W = 28 AB mag)
for each cluster according to the CATS models.
For different observed magnitudes or redshifts, 
the heights of the peaks change
as higher magnifications are more often required to yield brighter observed magnitudes.
Bottom: Corresponding plots for the Sharon (left) and Zitrin-LTM (right) models.
}
\end{figure*}

\section{High Redshift Predictions}
\label{sec:predictions}

In this section, we predict high-redshift number counts in the lensed and ``blank'' Frontier Fields.
For $z > 8$, we present ranges of predictions, as there are few observational constraints to date.
To predict lensed number counts, we convolved the field luminosity functions
with magnification histograms from lens models submitted by the community (\S\ref{sec:lensmodels}).

The numbers we quote are upper limits, as we do not address incompleteness here.
Foreground objects, including galaxies in the clusters and intracluster light,
will render a significant fraction of faint objects undetectable.
Incompleteness and contamination must both be quantified as a function of redshift and magnitude
typically by inserting simulated galaxies into the images and repeating one's analysis 
(searching for high-redshift candidates).

\subsection{Field Luminosity Functions}
\label{sec:fieldLF}

The UV luminosity function (LF) of star forming Lyman Break Galaxies (LBGs) in the field
has been fairly well constrained out to $z \sim 8$ where \~100 candidates have been discovered
\citep{Bouwens11b, Oesch12b, Yan12, Bradley12b, Dunlop13, McLure13, Schenker13, Schmidt14, Bouwens14}.
Estimating expected counts at $z \sim 9$ and greater requires extrapolation from lower redshifts
tempered with the handful of $z \gtrsim 9$ candidates discovered to date.


We adopted the best fit $z \sim 8$ LF from \cite{Bradley12b}.
They found the observed number counts as a function of luminosity were well fit by
a \cite{Schechter76} function $\phi(L) = \phi^* e^{-L/L^*} (-L/L^*)^\alpha$
with normalization $\phi^* = 4.3 \times 10^{-4}$ Mpc$^{-3}$,
characteristic rest-frame UV absolute magnitude $M^*_{UV} = -20.26$ AB,
and faint end slope $\alpha = -1.98$.
Other recent $z \sim 8$ studies \citep{McLure13, Schenker13, Schmidt14, Bouwens14} 
found similar results 
including similarly steep faint end slopes ($\alpha = -2.02$, $-1.94$, $-1.87$, and $-1.9$, respectively)
which help boost numbers of lensed faint galaxies (\S\ref{sec:lensLF}).
We explicitly tested the \cite{Schmidt14} LF and found it gives very similar expected number counts 
as those presented in \S\ref{sec:counts}.

For $z > 8$, we evolved the \cite{Bradley12b} LF by $dM^*/dz$ assuming $\phi^*$ and $\alpha$ remain constant.
\cite{Bouwens06} and most subsequent works found such $M^*(z)$ evolution to fit observed number counts well.
We adopted $dM^*/dz = 0.36$ and $dM^*/dz = 1.05$ as our fiducial ``optimistic'' and ``pessimistic'' cases, respectively.
The former is roughly consistent with the evolution observed at lower redshifts ($4 < z < 8$),
while the latter is consistent with the low numbers of $z \gtrsim 9$ candidates discovered to date
(\citealt{Oesch12a, Oesch13}; as discussed in \citealt{Coe13}).
However we acknowledge that the UV luminosity function evolution may be more accurately described
by evolution in the other parameters (\phistar\ and $\alpha$) or all three
as found more recently by \cite{Oesch14} and \cite{Bouwens14}.
%


We also considered the \cite{Bouwens12a} and \cite{Bouwens14}
evolving LFs derived from $4 < z < 8$ observations
in which all three parameters evolve with redshift.
These yield even more galaxies than the ``optimistic'' LF evolution we adopted above, 
especially for lensed $z > 8$ galaxies
as the faint end slope continues to steepen to $\alpha < -2$.
Bouwens et al.~show that 
extrapolating these relations to higher redshifts
significantly overestimates the numbers of
faint $z \sim 9$ and $z \sim 10$ galaxies observed in the field.
We adopted a fixed faint end slope $\alpha = -1.98$
so as not to exacerbate the disagreements with these results.
The Frontier Fields observations are required to better constrain faint $z > 8$ number counts
within our predicted ranges
and perhaps to constrain the minimum luminosity of high-redshift galaxies \citep{MashianLoeb13}.



\subsection{Lensed Luminosity Functions}
\label{sec:lensLF}

We estimated lensed number counts
by convolving the above luminosity functions 
with lens model magnification maps submitted by the community (\S\ref{sec:lensmodels}).
We performed this calculation in two ways, 
yielding expected numbers of lensed images and unique sources.

For the first method (yielding lensed images),
for every pixel within the WFC3/IR FOV (4.65 arcmin\squared),
we brightened the luminosity function $\phi(L)$ by the magnification $\mu$
and divided by the magnification to account for the reduced area.
We then summed over all pixels $i$ each with lensed area $A_{\rm pixel}$ in the image
to yield lensed number counts (``arcs'')
as a function of observed luminosity:
\begin{equation}
  N_{\rm arcs}(L_{\rm obs}) = \sum_{i}  \phi(\mu_i L)  \frac{ A_{\rm pixel} }{\mu_i}.
\end{equation}
Galaxies which are lensed to form multiple images are counted multiple times.
Thus observed catalogs can be compared directly to these predictions.


Our second estimation method yields numbers of unique source galaxies.
For those modelers who provided deflection maps (CATS, Sharon, Zitrin),
we delensed the magnification map back to the source plane sampled at higher resolution
(with pixels of area $A_s$) at the desired redshift.\footnote{
We began by rebinning the deflection field to derive values at the pixel corners (rather than their centers).
We then divided each magnification map pixel into two triangles, both assuming the magnification value in the square pixel.
We delensed the vertices of each triangle using our rebinned deflection field.
Our use of triangles ensures unique delensed regions in the source plane.
Squares could get twisted upon delensing.
Alternatively, we could have skipped the initial rebinning
and simply measured each triangle's magnification as the ratio of its lensed and delensed areas.
This yields nearly identical results.}
We identified all overlapping delensed areas and retained the highest magnifications at those pixels.
This corresponds to retaining only the brightest of each multiply-imaged galaxy.
Figure \ref{fig:sourceplanes} shows our result for the delensed CATS $z = 9$ magnification maps
as an illustrative example.

We then summed over all source pixels $j$, brightening the luminosity function by the magnifications
(but {\em not} reducing the areas since we are already working in the delensed source plane)
to yield lensed number counts of unique background galaxies:
\begin{equation}
  N_{\rm galaxies}(L_{\rm obs}) = \sum_{j}  \phi(\mu_j L)  A_{\rm s}.
\end{equation}



\subsection{Source Plane Search Volumes}
\label{sec:volumes}

Based on the delensed magnification maps from CATS, Sharon, and Zitrin, 
we can estimate the total source plane search areas as a function of magnification.
These search areas correspond to search volumes within, say, $\Delta z = 1$
of a given redshift as plotted for reference in Figure \ref{fig:volume} \citepeg{Hoggcosmo}.
%
%
In Figure \ref{fig:volumes}, we plot the cumulative search area
(and $z \sim 9$ volume)
as a function of magnification
for all six clusters according to the CATS and Zitrin-LTM models.
(T.~Johnson et al., 2014, in preparation, present a corresponding plot for the Sharon models.)
We also show the strong general agreement between the total lensed areas
according to CATS, Sharon, and Zitrin-LTM.
Based on these models, the full survey should yield 
\~5 arcmin\squared\ (\~9,000 Mpc\cubed\ at $z \sim 9$) 
of source plane search area in the 6 lensed WFC3/IR fields.
About 10\% of that should be magnified by a factor of 6 (\~2 magnitudes) or greater.
The blank parallel fields will yield about 5--6 times the search area of the lensed fields
(\~28 \sqarcmin, or \~50,000 Mpc\cubed\ at $z \sim 9$).
We stress these are upper limits as we have not accounted for search area lost due to foreground objects.

\subsection{Predicted Number Counts}
\label{sec:counts}

Figure \ref{fig:lensLFs} shows our ``optimistic'' $z \sim 9$ LF
(the \citealt{Bradley12b} LF with \Mstar\ fainter by 0.36 mag)  
lensed by every submitted lens model,
yielding estimated numbers of lensed images as a function of observed magnitude.
(For this plot, we count galaxies in the image plane allowing multiple images of galaxies to count multiple times.)

The variety of lens models yield broadly consistent predictions for lensed luminosity functions,
especially at fainter observed magnitudes ($> M^*$).
Some consistency is guaranteed by the faint end slope $\alpha \sim -2$.
This slope means that faint number counts are roughly inversely proportional to luminosity squared
($\phi \propto A L^{-2}$).
Thus for a given observed brightness,
a magnification $\mu$ decreases the search area $A$ by $\mu^{-2}$
but compensates by brightening fainter ($L / \mu$) galaxies which are more numerous by $\mu^2$
\citep{Broadhurst95}.

For our remaining calculations, we considered source galaxy counts,
discarding the fainter multiple images of any galaxy
as described in the previous subsections.
Figure \ref{fig:lensLFcats} shows our optimistic $z \sim 9$ LF
magnified by the delensed CATS, Sharon, and Zitrin-LTM models for each cluster.

We next considered the full survey and both the ``optimistic'' ($dM^*/dz = 0.36$) and ``pessimistic'' ($dM^*/dz = 1.05$)
scenarios discussed in \S\ref{sec:fieldLF}.
We plot cumulative expectations as a function of magnitude in Figure \ref{fig:FF6}.
We then plot the cumulative numbers for F160W $<$ 28.7 AB
as a function of redshift 
in the left panel of Figure \ref{fig:FF6cum}.
Integrating one more time, we plot total expected cumulative counts for F160W $<$ 28.7 AB as a function of redshift
in the right panel of Figure \ref{fig:FF6cum}.

We also tabulate some of these results in Table \ref{tab:counts}.
In the optimistic case, we expect $z > 9$ totals of \~40 lensed galaxies and \~30 field galaxies.
In the pessimistic case, we may only find \~9 lensed and \~3.5 field galaxies at $z > 9$.
Roughly, we can expect either six or one $z > 9$ candidate(s)
in each of the 12 Frontier Fields (lensed or blank)
depending on how smoothly (``optimistic'') or dramatically (``pessimistic'')
the LF declines at $z > 8$.
%
Again, we emphasize incompleteness may significantly reduce these numbers.

In Figure \ref{fig:FF6int}, we highlight the fact that
the lensed Frontier Fields are the deepest images ever obtained.
The UDF still has the greatest exposure time of any \HST\ image,
including 253 orbits WFC3/IR, 84 of which were obtained in F160W
yielding a 5\sig\ point source detection limit of 29.45 AB mag
\citep{Koekemoer13}.
The FF 24-orbit F160W images have 5\sig\ depths of \~28.7 AB mag
(0.75 mag fainter),
but a significant fraction (about half) of the source area 
will be lensed by factors of 2 (0.75 mag) or greater (Figure \ref{fig:volumes}).
Figure \ref{fig:FF6int} shows that significant numbers of galaxies intrinsically fainter than the UDF detection limit
should be detected out to $z \sim 8$
with perhaps a few detected out to $z \sim 10$ or 11 or so.

In fact, significant numbers of the detected galaxies will be magnified by factors up to \~10 (2.5 mag) or more.
In Figure \ref{fig:mobsint}, we plot (top left panel) the magnification likelihoods for $z = 9$ galaxies
as a function of observed magnitude as lensed by the Abell 2744 CATS model.
In the remaining panels,
we plot the $z = 9$ magnification likelihood distributions for all clusters according to multiple models
for a galaxy observed (lensed to) F160W = 28 AB mag.
Extreme magnifications ($\mu > 30$) are expected to be far less frequent.
In a way this is fortunate as such high magnification estimates for individual galaxies also have large uncertainties
\citepeg{Bradac09}.

%

\begin{deluxetable*}{llcccccc}
\tablecaption{\label{tab:obs}Observations Obtained to Date (April 2014)}
\tablehead{
\colhead{}&
\colhead{}&
\colhead{}&
\colhead{R.A.}&
\colhead{Decl.}&
\colhead{ACS}&
\colhead{WFC3/IR}&
\colhead{Spitzer}\\
\colhead{Cluster}&
\colhead{Field}&
\colhead{Name\supa}&
\colhead{(J2000)}&
\colhead{(J2000)}&
\colhead{(orbits)}&
\colhead{(orbits)}&
\colhead{(hours)}
}
\startdata
Abell 2744\supb & cluster & FFC1 & 00:14:21.20 & -30:23:50.1 & 16\supc & 71\supd & 100 \\
Abell 2744\supb & parallel & FFB1 &  00:13:53.27 & -30:22:47.8 & 87\supe & 0 & 100\\
MACSJ0416.1-2403\supf & cluster & FFC2 & 04:16:08.90 & -24:04:28.7 & 84\supg & 7\super{d,g} & 100\\
MACSJ0416.1-2403\supf & parallel & FFB2 & 04:16:33.40 & -24:06:49.1 & 0 & 80 & 100
\enddata
\tablenotetext{1}{Designation adopted here numbering observations and with C = Cluster; B = ``Blank'' parallel field}
\tablenotetext{2}{Frontier Fields \HST\ DD 13495 (PI Lotz), \SST\ 90257 (PI Soifer)}
\tablenotetext{3}{Includes observations from \HST\ GO 11689 (PI Dupke)} 
\tablenotetext{4}{Includes observations from \HST\ GO 11386 (PI Rodney)}
\tablenotetext{5}{Includes observations from \HST\ GO 13389 (PI Siana)}
\tablenotetext{6}{Frontier Fields \HST\ DD 13496 (PI Lotz), \SST\ 90258 (PI Soifer)}
\tablenotetext{7}{Includes observations from CLASH \HST\ GO 12459 (PI Postman)}
\end{deluxetable*}

\vspace{0.5in}
\section{Observations and Analysis}
\label{sec:obs}

We present high redshift candidates lensed by the first two Frontier Fields clusters
and in the first ``blank'' parallel field to be observed with WFC3/IR.
No Frontier Field has yet to be observed deeply (70 orbits) with both ACS and WFC3/IR,
but deep WFC3/IR imaging is available for Abell 2744 and the MACSJ0416.1-2403 parallel field
(as summarized in Table \ref{tab:obs}).
Deep Spitzer observations (100 hours) have also been obtained for the first two clusters and parallel fields.

The deep WFC3/IR and Spitzer imaging of Abell 2744 has been studied previously
\citep{Atek14,Laporte14,Zheng14}.
And \cite{Bradley13} presented brighter high-redshift candidates lensed by MACSJ0416.1-2403 
based on shallower 20-orbit, 16-band \HST\ imaging from CLASH.


\subsection{HST Images}
\label{sec:images}

We analyzed the version 1.0 \HST\ image mosaics produced by STScI
and available via the Mikulski Archive for Space Telescopes (MAST).\footnote{
\href{http://archive.stsci.edu/prepds/frontier/}
{http://archive.stsci.edu/prepds/frontier/}}
Associated README files describe in detail the observations and image processing,
which we summarize here.
The image mosaics incorporate
all images obtained to date in the Frontier Fields filters from multiple programs (see Table \ref{tab:obs}).

The STScI pipeline corrected all images for bias, dark current, and flat fields with up-to-date reference files.
Satellite trails were manually identified and masked.
WFC3/IR image processing included 
up-the-ramp cosmic ray rejection 
and manual masking of persistence and bad pixels.
The WFC3/IR images were also corrected for time-variable sky emission,
most significantly due to Helium line emission at 1.083\um\ affecting F105W observations
\citep{Brammer14ISR}.
%
This effect has been observed and addressed previously 
in CLASH \citep{Coe13}
and in the Ultra Deep Field \citep{Koekemoer13}.
ACS images were corrected for CTE (charge transfer efficiency) and bias striping.
Separate ``self-calibrated'' ACS images were also produced but not analyzed here.

The calibrated images were all aligned
to a common 0.06\arcsec\ pixel grid
using procedures from Astrodrizzle \citep{Gonzaga12}
and as outlined in \cite{Koekemoer02, Koekemoer11}.
Cosmic rays were identified in the multiple ACS exposures and rejected.
Inverse variance maps (IVMs) were produced quantifying the uncertainty in each pixel
before accounting for 
correlated pixel noise (an additional 10-15\%)
and Poisson source noise.
These IVMs were used as weights to drizzle-combine the images obtained in each filter.
The final seven filter image mosaics are aligned with one another to precisions of 0.002--0.005\arcsec\ RMS.


\subsection{Photometric Redshifts}
\label{sec:photoz}

Based on the STScI version 1.0 seven-filter \HST\ image mosaics and IVMs of each field,
we measured photometry and photometric redshifts in a manner similar to our analyses of CLASH images
\citepeg{Coe13,Jouvel14}.
We created a weighted sum of all four \HST\ WFC3/IR images to be used as our detection image for each field.
And we took the inverse square root of each inverse variance map to create an RMS map for each filter.

We ran SExtractor \citep{SExtractor} version 2.8.6 in dual-image mode
to detect objects in each field
and define their isophotal apertures for photometry to be measured in each filter image.
%
For object detection, we required 9 contiguous pixels in the detection image above the RMS 
measured after background subtraction in $3\times3$ grids of $64\times64$ pixels.
We set the deblending of adjacent objects to 64 levels of 0.0001 minimum contrast.
For photometry, 
we used a 24-pixel-wide rectangular annulus around each object 
to estimate and subtract the local background.

We estimated Bayesian photometric redshifts using BPZ \citep{Benitez00,Coe06},
one of the top performing methods in controlled tests \citep{Hildebrandt10}.
We used updated spectral models and priors (Ben\'itez et al.~in prep.) as described in \cite{Coe13}.
Briefly, the model spectral energy distributions (SEDs) are
originally from PEGASE \citep{Fioc97}
but recalibrated based on observed photometry and spectroscopic redshifts from FIREWORKS \citep{Wuyts08}.
The templates were selected to encompass ranges of metallicities, extinctions, and star formation histories
observed for the vast majority of real galaxies.

BPZ fits our observed photometry to these redshifted models
and then tempers the \chisq\ redshift likelihood with 
a Bayesian prior $P(z,T|m)$ on each galaxy redshift and type given its observed magnitude.
The updated prior is based on galaxy samples observed in COSMOS \citep{Ilbert09},
GOODS-MUSIC \citep{Grazian06, Santini09}, and the UDF \citep{Coe06}.

This prior may be conservative in that it assumes 
high-redshift ($z > 6$) galaxies are less likely 
than their primary interlopers (red $z \sim 2$ galaxies).
\cite{Bouwens12c} argued that at these faint magnitudes,
red $z \sim 2$ galaxies are actually less common \citep{Giallongo05}
than high redshift galaxies \citep{Bouwens11b}.
Our conservative prior may be more appropriate at $z > 8$
given the declines in number counts suggested by field searches
(most recently \citealt{Bouwens14}).
For our high-redshift searches,
we consider both the conservative prior
as well as a flat prior which assumes all redshifts are equally likely
(the ``maximum likelihood'' results output by BPZ).


\begin{deluxetable*}{lccrrrrrrr}
\tablecaption{\label{tab:FFC1}Best high redshift ($z > 7.5$) candidates lensed by Abell 2744 (FFC1) based on deep WFC3/IR and shallow ACS imaging}
\tablehead{
\colhead{}&
\colhead{}&
\colhead{}&
\colhead{F606W\supc}&
\colhead{F814W\supc}&
\colhead{F105W\supc}&
\colhead{F125W\supc}&
\colhead{F140W\supc}&
\colhead{F160W\supc}&
\colhead{}\\
\colhead{ID\supa}&
\colhead{$z_b$\supb}&
\colhead{$P(z>4)$\supd}&
\colhead{(nJy)}&
\colhead{(nJy)}&
\colhead{(nJy)}&
\colhead{(nJy)}&
\colhead{(nJy)}&
\colhead{(nJy)}&
\colhead{magnif.\supd}
}
\startdata
FFC1-2493-2561\suph&
$8.2 \range{1.8}{8.4}$&
0.95&
$11.1 \pm 3.7$&
$1.5 \pm 5.5$&
$14.8 \pm 2.4$&
$55.2 \pm 3.4$&
$68.5 \pm 3.3$&
$74.3 \pm 3.0$&
$1.7 \range{1.5}{2.4} \range{1.4}{2.7}$\\
FFC1-2306-3090\supf\suph&
$8.1 \range{7.8}{8.2}$&
1.00&
$-6.8 \pm 3.8$&
$4.7 \pm 5.6$&
$25.2 \pm 2.3$&
$69.4 \pm 3.4$&
$77.1 \pm 3.2$&
$68.8 \pm 2.9$&
$2.6 \range{2.1}{10} \range{2.1}{7.4}$\\
FFC1-2136-2432\suph&
$7.9 \range{7.2}{8.1}$&
0.98&
$1.4 \pm 2.0$&
$1.2 \pm 2.9$&
$10.9 \pm 1.3$&
$27.4 \pm 1.9$&
$23.8 \pm 1.8$&
$26.1 \pm 1.7$&
$2.1 \range{1.9}{2.4} \range{1.9}{2.4}$\\
FFC1-2508-2497\supf\supg\suph&
$7.9 \range{7.7}{8.1}$&
0.99&
$12.3 \pm 4.6$&
$19.6 \pm 6.7$&
$53.1 \pm 3.6$&
$143.8 \pm 5.1$&
$145.0 \pm 5.2$&
$159.0 \pm 4.7$&
$1.6 \range{1.4}{2.3} \range{1.3}{3.3}$\\
FFC1-2481-2561\suph&
$7.8 \range{7.5}{8.0}$&
0.99&
$0.6 \pm 4.8$&
$3.9 \pm 7.2$&
$40.4 \pm 3.1$&
$96.3 \pm 4.5$&
$117.9 \pm 4.3$&
$115.2 \pm 4.0$&
$1.7 \range{1.5}{2.4} \range{1.4}{2.7}$\\
FFC1-2555-2516\supf\suph&
$7.6 \range{1.2}{7.8}$&
0.83&
$6.8 \pm 4.4$&
$-15.4 \pm 6.5$&
$37.0 \pm 2.8$&
$75.2 \pm 4.0$&
$82.1 \pm 3.9$&
$86.5 \pm 3.5$&
$1.6 \range{1.4}{2.1} \range{1.1}{2.1}$\\
FFC1-2355-2566\suph&
$1.1 \range{0.6}{7.7}$&
0.58&
$-5.2 \pm 2.4$&
$-0.8 \pm 3.6$&
$11.4 \pm 1.6$&
$19.4 \pm 2.3$&
$18.0 \pm 2.2$&
$17.8 \pm 2.0$&
$1.8 \range{1.7}{3.3} \range{1.6}{3.1}$
\\
\vspace{-0.1in}
\enddata
\tablenotetext{1}{IDs encode field names (Table \ref{tab:obs}) followed by the last 4 significant digits of R.A.~and Decl.~for each object. For example, FFC1-2493-2561 is located at R.A.~(J2000) = 00:14:{\bf 24.93}; Decl.~(J2000) = -30:2{\bf 2:56.1}.}
\tablenotetext{2}{Bayesian photometric redshift [95\% C.L.].}
\tablenotetext{3}{Measured fluxes are given in nJy. AB magnitudes are given by $m_{AB} = 31.4 - 2.5 \log_{10}(F_{\rm nJy})$.}
\tablenotetext{4}{Estimated Bayesian probability of being at high redshift ($z > 4$) with a conservative prior.}
\tablenotetext{5}{Magnification estimates based on submitted models. We quote the median and 68.3\% range of parametric models in brackets followed by the 68.3\% range of all models in brackets.}
\tablenotetext{6}{Also reported by \cite{Atek14}.}
\tablenotetext{7}{Studied further by \cite{Laporte14}.}
\tablenotetext{8}{Also reported by \cite{Zheng14}.}
\end{deluxetable*}

\begin{figure}
\centerline{
\includegraphics[width = 0.49\textwidth]{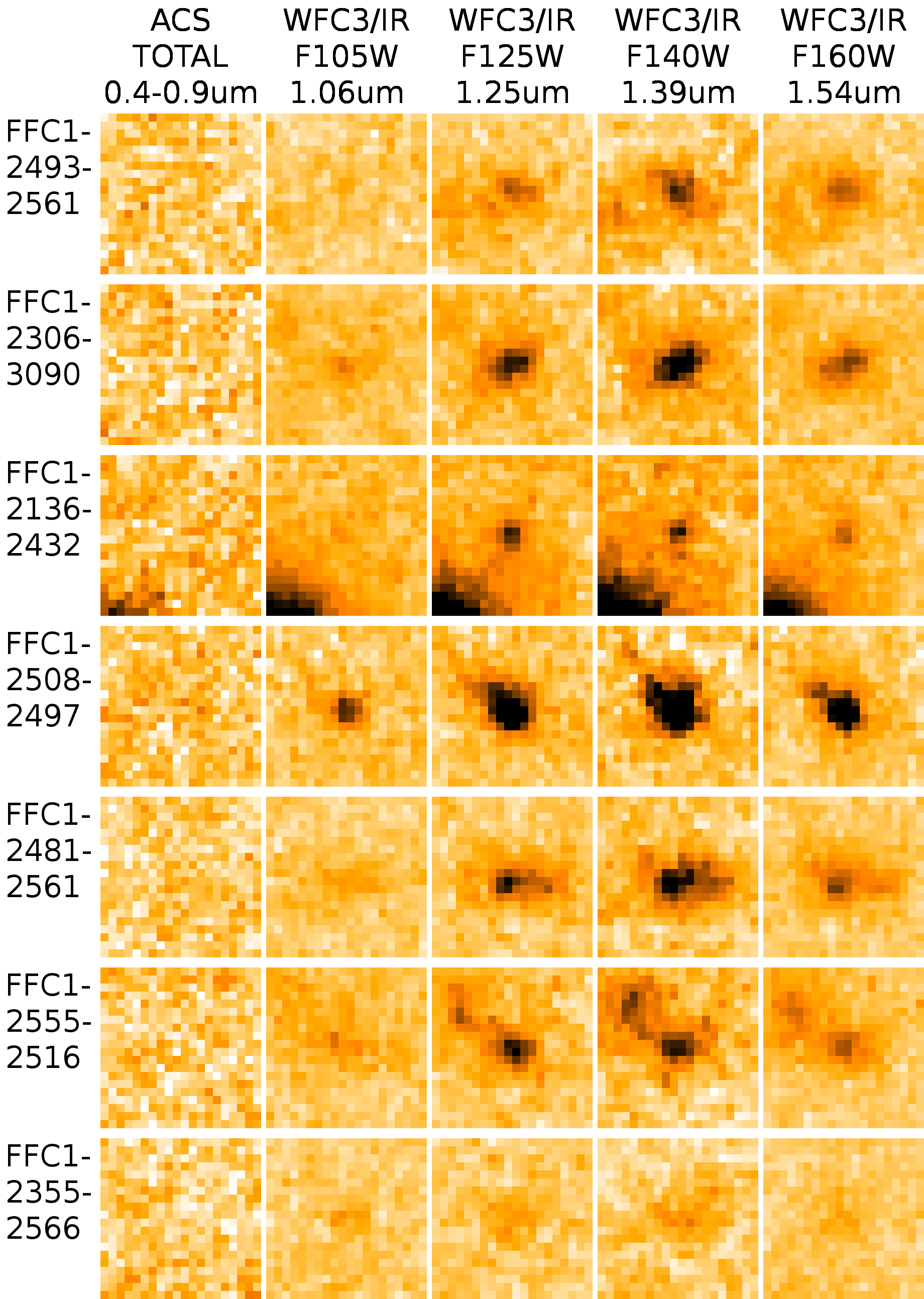}
}
\caption{
\label{fig:FFC1stamps}
Best $z > 7.5$ candidates lensed by Abell 2744 (see Table \ref{tab:FFC1}).
Image stamps are $1.2'' \times 1.2''$.
}
\end{figure}

\begin{figure*}
\centerline{
\includegraphics[width = 0.49\textwidth]{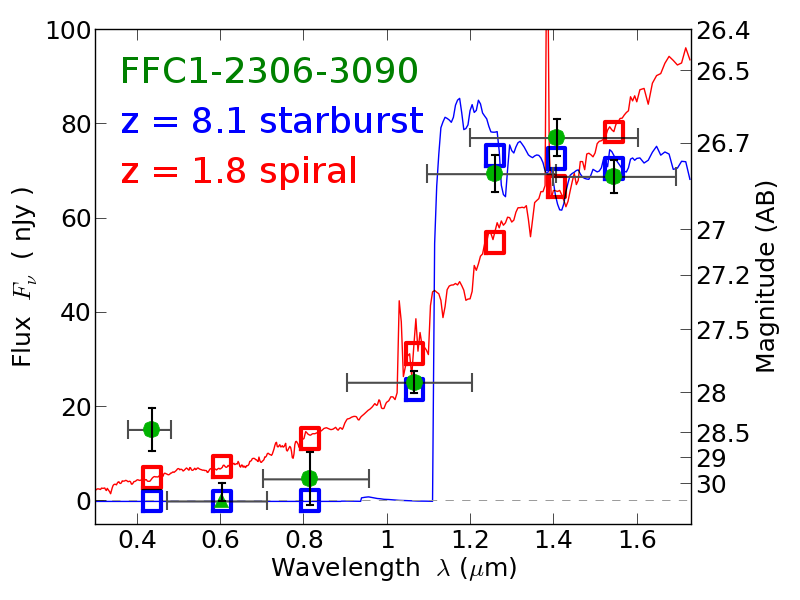}
\includegraphics[width = 0.49\textwidth]{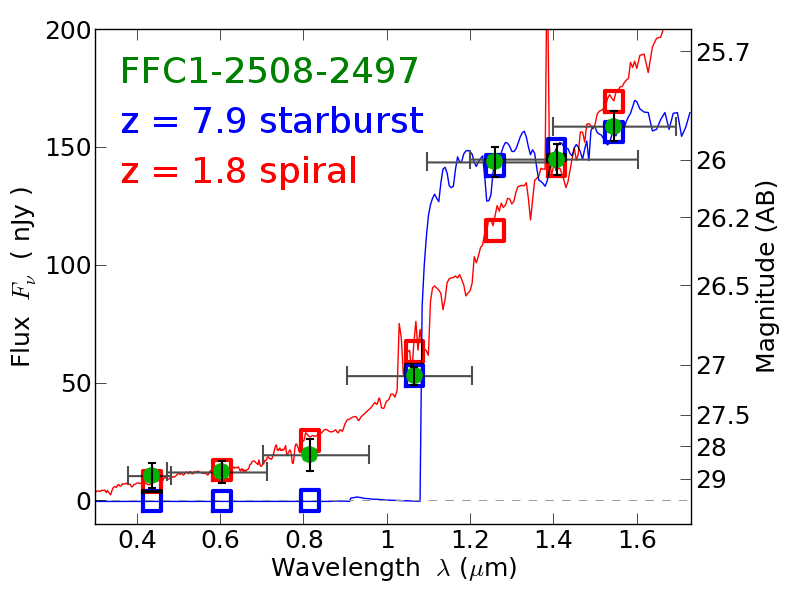}
}
\caption{
\label{fig:seds_FFC1}
SEDs of two confident $z \sim 8$ candidates lensed by Abell 2744
based on deep WFC3/IR and shallow ACS imaging
(see Table \ref{tab:FFC1}).
We plot the observed fluxes as filled green circles (or triangles for non-detections)
with 1\sig\ vertical error bars.
Horizontal error bars mark the filter wavelength ranges.
We plot our best fitting spectral template in blue
as well as the best $z < 4$ template in red.
We integrate these spectra through the \HST\ filters
to plot model fluxes as open squares.
Both candidates were also identified by \cite{Atek14} and \cite{Zheng14},
and FFC1-2508-2497 (right) was studied further by \cite{Laporte14}.
Deep ACS imaging is upcoming for these candidates.
}
\end{figure*}


\begin{deluxetable*}{lccccrrrr}
\tablecaption{\label{tab:FFB2}Best high redshift ($z > 7.5$) candidates in FFB2, the blank field adjacent to MACSJ0416.1-2403, based on deep WFC3/IR imaging}
\tablewidth{\textwidth}
\tablehead{
\colhead{}&
\colhead{R.A.}&
\colhead{Decl.}&
\colhead{}&
\colhead{}&
\colhead{F105W\supc}&
\colhead{F125W\supc}&
\colhead{F140W\supc}&
\colhead{F160W\supc}\\
\colhead{ID\supa}&
\colhead{(J2000)}&
\colhead{(J2000)}&
\colhead{$z_b$\supb}&
\colhead{$P(z>4)$\supd}&
\colhead{(nJy)}&
\colhead{(nJy)}&
\colhead{(nJy)}&
\colhead{(nJy)}
}
\startdata
FFB2-3642-6304&
04:16:36.42&
-24:06:30.4&
$8.7 \range{1.7}{9.2}$&
0.85&
$-1.5 \pm 0.8$&
$8.2 \pm 1.3$&
$9.0 \pm 1.1$&
$9.9 \pm 1.1$\\
FFB2-3597-6481&
04:16:35.97&
-24:06:48.1&
$8.6 \range{1.7}{9.2}$&
0.79&
$2.1 \pm 1.2$&
$14.2 \pm 1.8$&
$20.8 \pm 1.6$&
$21.7 \pm 1.5$\\
FFB2-3043-6011&
04:16:30.43&
-24:06:01.1&
$8.1 \range{1.2}{8.4}$&
0.73&
$4.6 \pm 1.1$&
$14.3 \pm 1.7$&
$16.1 \pm 1.5$&
$13.5 \pm 1.4$\\
FFB2-3157-5248&
04:16:31.57&
-24:05:24.8&
$7.9 \range{0.8}{8.2}$&
0.44&
$9.6 \pm 2.2$&
$29.8 \pm 3.4$&
$21.8 \pm 3.0$&
$25.1 \pm 2.8$\\
FFB2-3045-6002&
04:16:30.45&
-24:06:00.2&
$7.9 \range{1.2}{8.2}$&
0.85&
$7.7 \pm 1.3$&
$20.0 \pm 1.9$&
$18.4 \pm 1.7$&
$16.5 \pm 1.6$
\\
\vspace{-0.1in}
\enddata
\tablenotetext{1}{IDs encode field names (Table \ref{tab:obs}) followed by the last 4 significant digits of R.A.~and Decl.~for each object.}
\tablenotetext{2}{Bayesian photometric redshift [95\% C.L.].}
\tablenotetext{3}{Measured fluxes are given in nJy. AB magnitudes are given by $m_{AB} = 31.4 - 2.5 \log_{10}(F_{\rm nJy})$.}
\tablenotetext{4}{Estimated Bayesian probability of being at high redshift ($z > 4$) with a conservative prior.}
\end{deluxetable*}

\begin{figure}
\centerline{
\includegraphics[width = 0.49\textwidth]{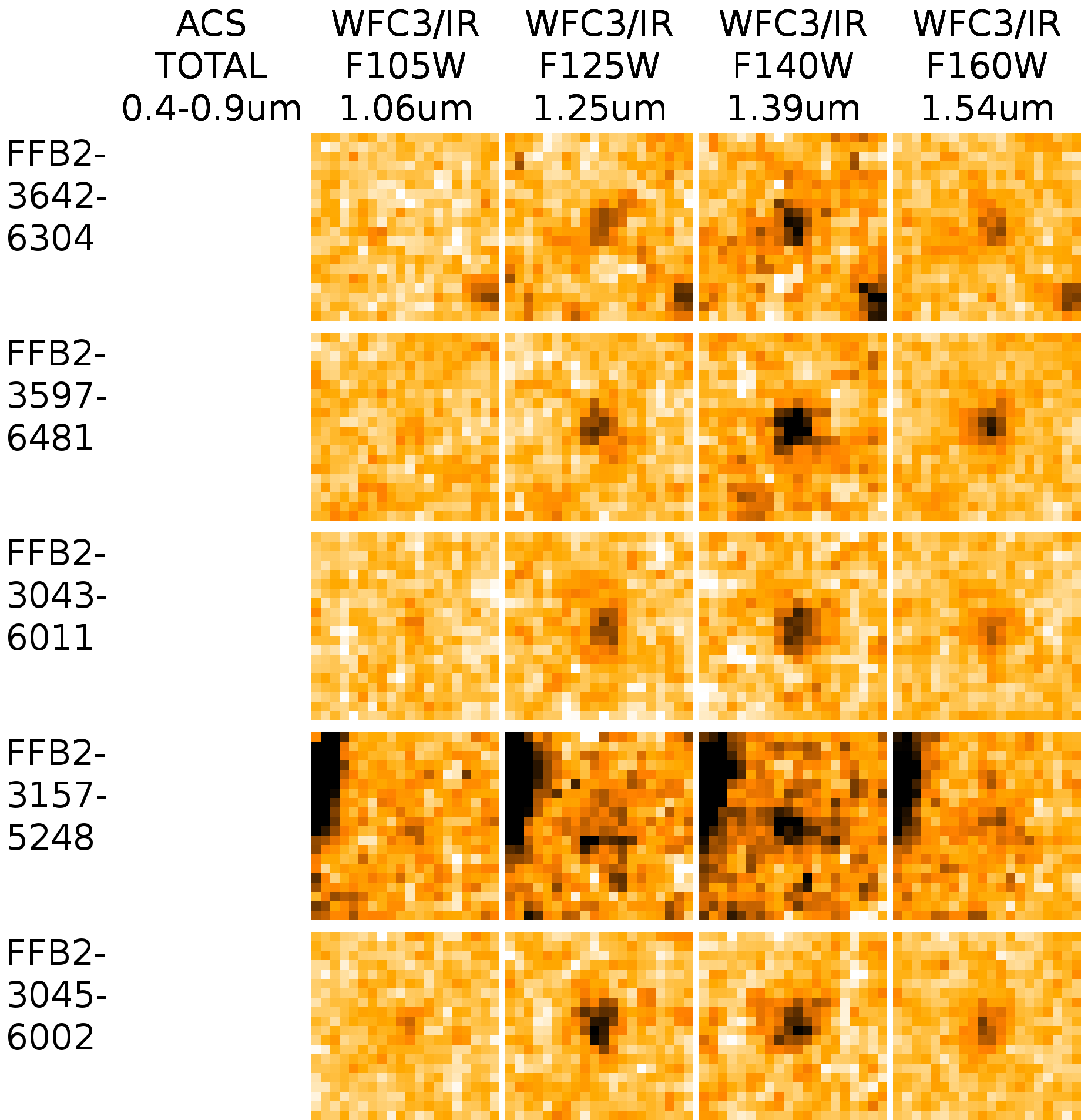}
}
\caption{
\label{fig:FFB2stamps}
Best $z > 7.5$ candidates in FFB2
(see Table \ref{tab:FFB2}).
Image stamps are $1.2'' \times 1.2''$.
}
\end{figure}

\begin{figure}
\centerline{
\includegraphics[width = 0.49\textwidth]{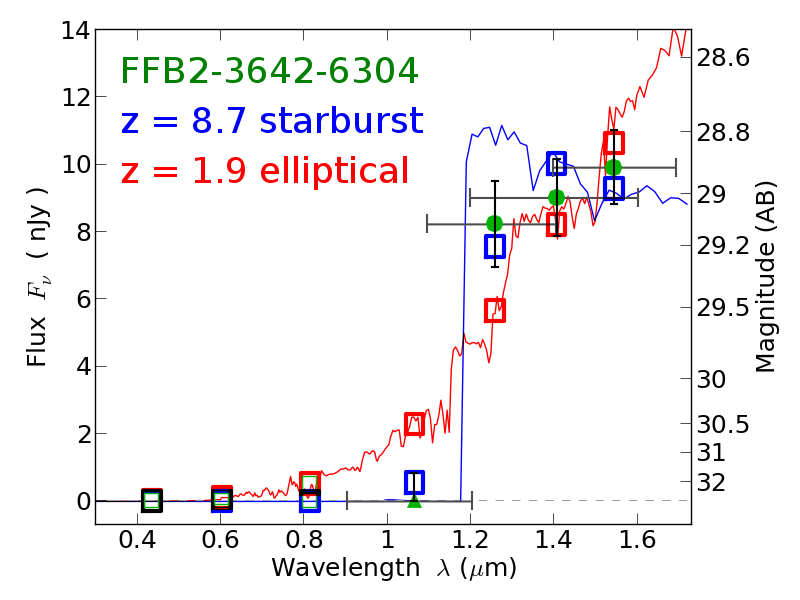}
}
\caption{
\label{fig:seds_FFB2}
SED of a $z \sim 9$ candidate in FFB2
(see Table \ref{tab:FFB2}),
plotted as in Figure \ref{fig:seds_FFC1}.
Even without ACS imaging, 
we can identify $z \sim 9$ candidates as they drop out of the WFC3/IR F105W filter.
}
\end{figure}

\begin{deluxetable*}{lccrrrrrrr}
\tablecaption{\label{tab:FFC2}Best high redshift ($z > 5.5$) candidates lensed by MACSJ0416.1-2403 (FFC2) based on deep ACS and shallow WFC3/IR imaging}
\tablehead{
\colhead{}&
\colhead{}&
\colhead{}&
\colhead{F606W\supd}&
\colhead{F814W\supd}&
\colhead{F105W\supd}&
\colhead{F125W\supd}&
\colhead{F140W\supd}&
\colhead{F160W\supd}&
\colhead{}\\
\colhead{ID\supa}&
\colhead{$z_b$\supb}&
\colhead{$P(z>4)$\supc}&
\colhead{(nJy)}&
\colhead{(nJy)}&
\colhead{(nJy)}&
\colhead{(nJy)}&
\colhead{(nJy)}&
\colhead{(nJy)}&
\colhead{magnif.\supe}
}
\startdata
FFC2-1153-4532\supf&
$2.1 \range{1.8}{9.3}$&
0.38&
$-0.2 \pm 1.3$&
$0.2 \pm 1.0$&
$3.1 \pm 6.0$&
$31.0 \pm 6.9$&
$46.6 \pm 7.5$&
$52.0 \pm 6.8$&
$1.8 \range{1.6}{2.6} \range{1.6}{5.7}$\\
FFC2-1151-4540&
$8.1 \range{7.3}{8.6}$&
1.00&
$-1.1 \pm 1.1$&
$-0.4 \pm 0.8$&
$9.2 \pm 5.1$&
$40.4 \pm 5.7$&
$40.4 \pm 6.2$&
$37.0 \pm 5.7$&
$1.8 \range{1.6}{2.6} \range{1.6}{5.7}$\\
FFC2-1148-4581&
$6.9 \range{1.9}{7.7}$&
0.95&
$-2.5 \pm 1.1$&
$-1.4 \pm 0.8$&
$20.7 \pm 4.9$&
$20.5 \pm 5.6$&
$15.2 \pm 6.4$&
$26.2 \pm 5.4$&
$1.7 \range{1.5}{2.5} \range{1.5}{5.5}$\\
FFC2-0744-5264&
$6.8 \range{1.1}{7.9}$&
0.63&
$-0.5 \pm 1.1$&
$-0.7 \pm 0.8$&
$13.4 \pm 5.1$&
$16.2 \pm 5.7$&
$10.7 \pm 6.0$&
$8.0 \pm 5.7$&
$3.7 \range{2.0}{6.1} \range{1.6}{5.3}$\\
FFC2-0960-3426\supg&
$6.3 \range{0.9}{6.4}$&
0.84&
$2.6 \pm 1.3$&
$8.2 \pm 1.0$&
$42.4 \pm 6.3$&
$30.4 \pm 7.2$&
$42.1 \pm 7.8$&
$30.7 \pm 7.1$&
$15 \range{8.7}{249} \range{6.2}{142}$\\
FFC2-0500-3510\supg&
$6.2 \range{0.8}{6.3}$&
0.80&
$-0.2 \pm 1.2$&
$7.6 \pm 0.9$&
$33.0 \pm 6.6$&
$24.9 \pm 6.6$&
$23.0 \pm 6.4$&
$31.5 \pm 6.6$&
$1.7 \range{1.4}{2.5} \range{1.2}{3.9}$\\
FFC2-1148-3435\supg&
$6.2 \range{5.7}{6.3}$&
0.99&
$2.4 \pm 1.6$&
$17.1 \pm 1.1$&
$66.4 \pm 7.3$&
$60.1 \pm 8.6$&
$49.7 \pm 9.6$&
$62.2 \pm 8.3$&
$23 \range{13}{29} \range{7.5}{34}$\\
FFC2-0900-3404&
$6.1 \range{0.9}{6.2}$&
0.94&
$-1.8 \pm 1.3$&
$10.6 \pm 1.0$&
$38.0 \pm 6.0$&
$32.8 \pm 6.8$&
$37.7 \pm 7.5$&
$33.6 \pm 6.8$&
$6.5 \range{3.9}{11} \range{3.4}{18}$\\
FFC2-1221-3595&
$6.1 \range{5.4}{6.2}$&
0.98&
$-1.2 \pm 1.1$&
$8.2 \pm 0.8$&
$30.8 \pm 5.0$&
$22.3 \pm 5.7$&
$29.8 \pm 6.3$&
$17.9 \pm 5.6$&
$4.2 \range{3.2}{9.3} \range{2.5}{21}$\\
FFC2-0829-5229&
$5.9 \range{0.5}{6.2}$&
0.68&
$-1.0 \pm 0.8$&
$4.1 \pm 0.6$&
$16.3 \pm 3.8$&
$6.7 \pm 4.3$&
$6.1 \pm 4.4$&
$12.7 \pm 4.3$&
$2.8 \range{1.6}{4.3} \range{2.1}{8.1}$\\
FFC2-0972-5201&
$5.9 \range{0.8}{6.2}$&
0.91&
$-0.4 \pm 1.4$&
$12.7 \pm 1.0$&
$36.4 \pm 6.6$&
$48.1 \pm 7.4$&
$46.0 \pm 8.0$&
$39.9 \pm 7.5$&
$1.9 \range{1.5}{2.7} \range{1.4}{3.4}$\\
FFC2-1321-4063&
$5.5 \range{4.9}{5.7}$&
1.00&
$-0.2 \pm 1.3$&
$25.4 \pm 1.0$&
$40.5 \pm 6.1$&
$34.4 \pm 7.1$&
$39.4 \pm 7.9$&
$31.2 \pm 6.9$&
$2.0 \range{1.7}{3.3} \range{1.6}{3.1}$\\
FFC2-0640-3143&
$5.5 \range{4.8}{5.7}$&
1.00&
$-0.7 \pm 0.8$&
$14.3 \pm 0.6$&
$21.3 \pm 4.1$&
$17.9 \pm 8.1$&
\nodata&
$18.7 \pm 4.7$&
$1.4 \range{1.4}{1.9} \range{0.7}{1.8}$
\\
\vspace{-0.1in}
\enddata
\tablenotetext{1}{IDs encode field names (Table \ref{tab:obs}) followed by the last 4 significant digits of R.A.~and Decl.~for each object. For example, FFC2-1153-4532 is located at R.A.~(J2000) = 04:16:{\bf 11.53}; Decl.~(J2000) = -24:0{\bf 4:53.2}.}
\tablenotetext{2}{Bayesian photometric redshift (95\% C.L.).}
\tablenotetext{3}{Measured fluxes are given in nJy. AB magnitudes are given by $m_{AB} = 31.4 - 2.5 \log_{10}(F_{\rm nJy})$.}
\tablenotetext{4}{Estimated Bayesian probability of being at high redshift ($z > 4$) with a conservative prior.}
\tablenotetext{5}{Magnification estimates based on submitted models. We quote the median and 68.3\% range of parametric models followed by the (usually expanded) additional 68.3\% range of all models.}
\tablenotetext{6}{Included as part of a search for lower likelihood $z > 7.5$ candidates.}
\tablenotetext{7}{Reported by \cite{Bradley13} based on CLASH imaging.}
\end{deluxetable*}

\begin{figure}
\centerline{
\includegraphics[width = 0.49\textwidth]{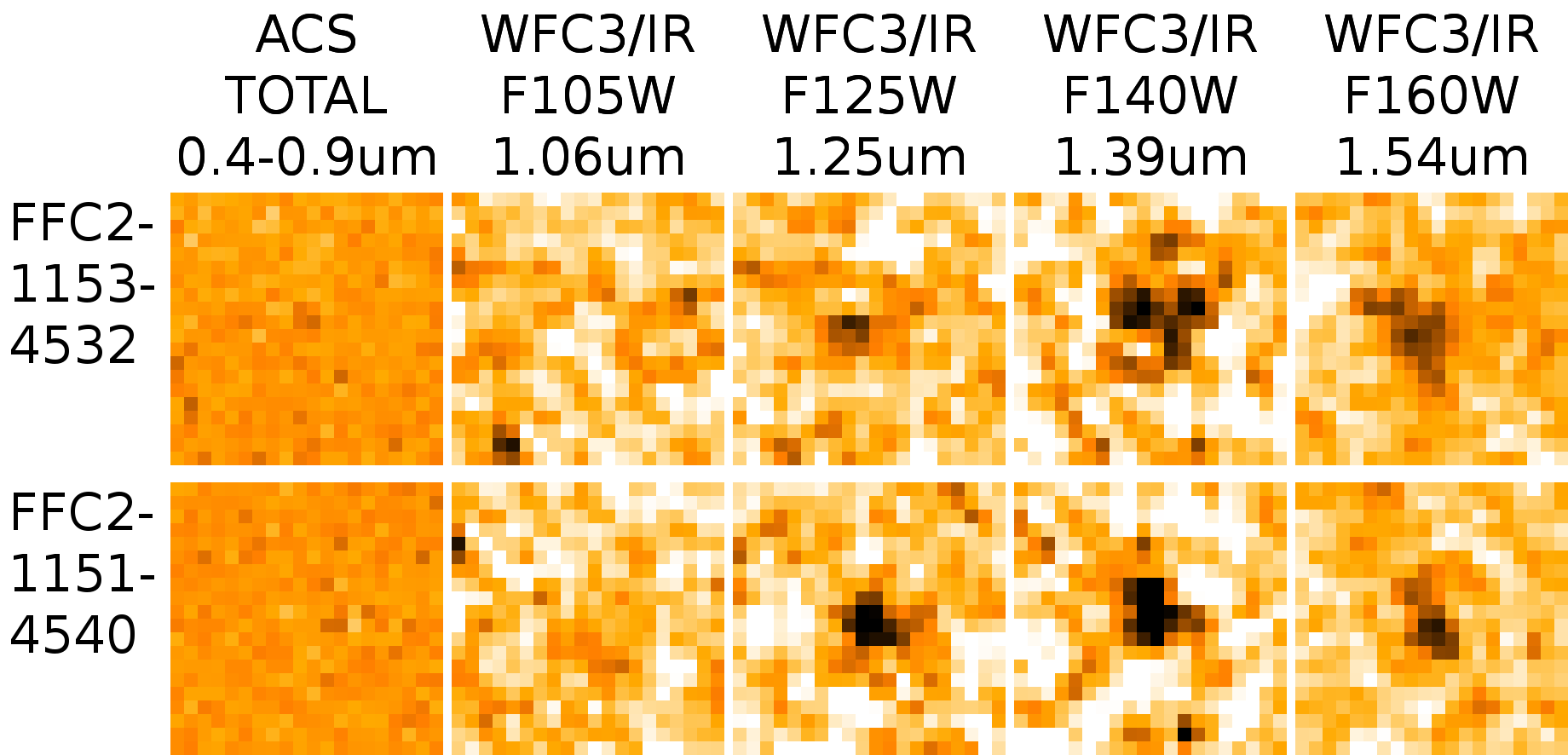}
}
\caption{
\label{fig:FFC2stamps}
Candidate $z > 7.5$ galaxies lensed by MACSJ0416.1-2403 (see Table \ref{tab:FFC2}).
Image stamps are $1.2'' \times 1.2''$.
}
\end{figure}

\begin{figure*}
\centerline{
\includegraphics[width = 0.49\textwidth]{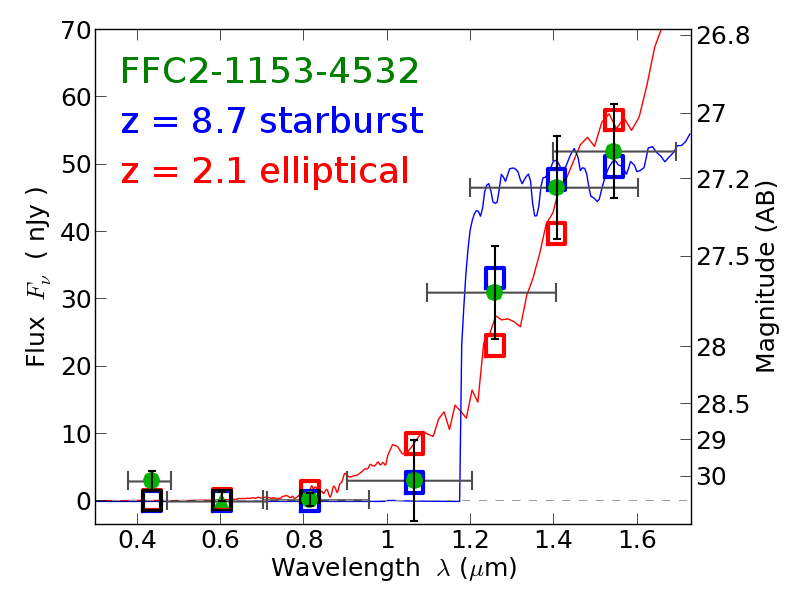}
\includegraphics[width = 0.49\textwidth]{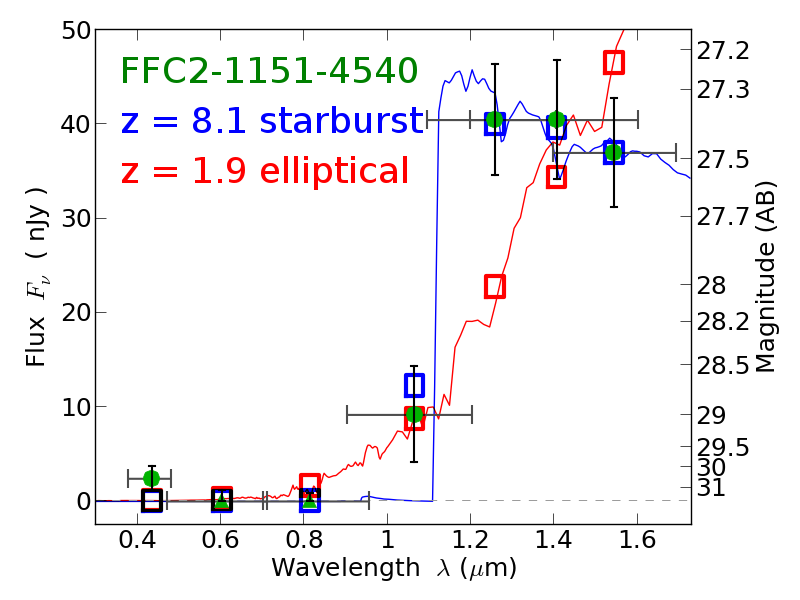}
}
\caption{
\label{fig:seds_FFC2}
SEDs of a possible $z \sim 9$ candidate (left) 
and a more confident $z \sim 8$ candidate (right)  
lensed by MACSJ0416.1-2403
based on deep ACS and shallow WFC3/IR imaging (see Table \ref{tab:FFC2})
plotted as in Figure \ref{fig:seds_FFC1}.
}
\end{figure*}

\section{High Redshift Candidates}
\label{sec:highz}

\subsection{Lensed by Abell 2744}
\label{sec:abell2744}

In Table \ref{tab:FFC1}, 
we present our most likely $z > 7.5$ candidates lensed by Abell 2744
as observed in the deep WFC3/IR imaging and shallow archival ACS imaging
(Figure \ref{fig:FFC1stamps}).
\cite{Atek14} previously noted three of these seven candidates,
and \cite{Laporte14} further analyzed the brightest $z \sim 8$ candidate FFC1-2508-2497.
\cite{Zheng14} reported all of these $z \sim 8$ candidates plus others not included here
due to our more conservative priors.
The upcoming deep ACS imaging should help discriminate between the $z \sim 2$ and $z \sim 8$ solutions for these objects.

We provide the estimated likelihood that each object is at high redshift ($z > 4$) versus low redshift
based on the Bayesian analysis including the conservative prior.
Here we only include candidates more likely to be high redshift than not,
discarding a number of candidates with conservative $P(z>4) < 0.15$.
In Figure \ref{fig:seds_FFC1}, 
we plot example SEDs of two of the more confident candidates, including FFC1-2508-2497.

Intriguingly, all of our $z \sim 8$ candidates are located in one corner of the image.
As noted by \cite{Zheng14}, this could be the densest concentration of $z \sim 8$ candidates yet observed.
\cite{Trenti12} discovered five $z \sim 8$ candidates within a diameter of 140\arcsec\ (6 comoving Mpc) in one BoRG field.
\cite{Zheng14} find nine $7 < z < 9$ candidates within a 20\arcsec\ search diameter.
We find at least four of these to be robust (conservatively)
and that all seven of our candidates fit within 110\arcsec.

According to the submitted lens models, magnifications in this region are relatively low, around a factor of 2.
In Table \ref{tab:FFC1}, we provide magnification estimates and uncertainties 
based on the ranges of models submitted by each team (altogether \~1,000 models per cluster).
We weight each modeler (CATS, Sharon, Zitrin, Williams, Bradac, Merten)
equally regardless of the number of models submitted.

We first provide the median and 68\% confidence range
for the methods which make use of the prior that light traces mass (CATS, Sharon, Zitrin).
We then quote the range allowed when including also the more flexible models 
which make no assumptions about light tracing mass (Williams, Bradac, Merten).
For example, FFC1-2508-2497 is likely magnified by a factor of \~1.6, or 1.4 -- 2.2 with 68\% confidence,
according to the light-traces-mass methods.
This range expands to 1.2 -- 3.2 when considering all methods.
For some other objects, the 68\% confidence range shrinks when all models are included.

\subsection{MACSJ0416.1-2403 parallel blank field}
\label{sec:macs0416par}

In Table \ref{tab:FFB2}, we report for the first time high-redshift ($z > 7.5$) candidates 
in the deep WFC3/IR imaging of FFB2, the blank field observed in parallel with MACSJ0416.1-2403.
Though no ACS imaging is available, 
$z > 7.5$ galaxies drop out significantly from F105W (Figure \ref{fig:filters}),
so we may already begin to search for candidates at these high redshifts.
We find higher redshift candidates in this field, out to $z \sim 8.7$.
We show WFC3/IR image stamps in Figure \ref{fig:FFB2stamps}
and an example SED in Figure \ref{fig:seds_FFB2}.
All of these candidates allow for lower redshift ($z < 4$) solutions within their 95\% confidence limits
again using our conservative prior.

\subsection{Lensed by MACSJ0416.1-2403}
\label{sec:macs0416}

In Table \ref{tab:FFC2}, we report new high-redshift candidates lensed by MACSJ0416.1-2403.
Due to the lack of deep WFC3/IR imaging, we do not yet expect to find many $z > 7.5$ candidates.
But we can search for brighter $z > 7.5$ candidates in the 6-orbit CLASH WFC3/IR imaging.
We find two separated by 1\arcsec:
a confident $z \sim 8$ candidate
and a less confident $z \sim 9$ candidate.
We present image stamps in Figure \ref{fig:FFC2stamps} and SEDs in Figure \ref{fig:seds_FFC2}

We also expand our search to $z > 5.5$
allowing us to compare with \cite{Bradley13}
who identified $5.5 < z < 8.5$ candidates in CLASH imaging of 18 clusters including MACSJ0416.1-2403.
They identified three strong candidates and four less confident candidates (with uncertainties extending to $z < 4$)
lensed by MACSJ0416.1-2403.
(They also noted one candidate likely to be a star, which we exclude here.)
Based on the deeper ACS imaging, two of their less confident candidates (\#546, \#419) are now more confidently high redshift.
One of their confident candidates (\#427) remains confident.
Another (\#1821) is surprisingly now more likely low redshift ($P(z>4) = 0.30$).
The third (\#1957) is not properly deblended in our detection, 
lowering its redshift estimate to $z \sim 5$.
We defer further analysis to the full ACS + WFC3/IR dataset.












\section{Discussion}
\label{sec:discussion}
\label{sec:conclusions}

The Frontier Fields program promises to yield significant numbers of high redshift candidates,
including \~10 -- 70 candidates at $z > 9$
where only 10 or so are known to date \citep{Zheng12N,Coe13,Ellis13,Oesch13,Oesch14}.
Importantly, the multiple sightlines will reduce uncertainties due to cosmic variance, or
chance observations of overdense or underdense fields.
The lensed fields are yielding nJy sources, the faintest galaxies observed to date.
All of the lens models agree roughly on the number of faint galaxies expected,
especially if the luminosity function faint end slope is $\alpha \sim -2$ as observed at $z \sim 8$.

In early Frontier Fields observations, including two deep (70 orbit) WFC3/IR images
(with deep ACS imaging yet to come),
we find no strong $z > 9$ candidates.
This contrasts with our expectation of \~6 $z > 9$ candidates per field
assuming smooth luminosity function evolution 
as observed from $z \sim 4$ to 8 continuing to $z > 9$.

Given the significant cosmic variance (more than $\pm 50\%$ per field; see \citealt{Bouwens14})
and possible strong clustering expected \citep{Trenti12},
the six independent sightlines of blank and lensed fields
observed by the full Frontier Fields program
will be key to constraining galaxy evolution in the first 600 million years.

\acknowledgements{
This work is supported by HST archival research program AR-13236.
Support for AZ is provided by NASA through Hubble Fellowship grant HST-HF-51334.01-A awarded by STScI.

The Frontier Fields are Director's Discretionary programs carried out with
the NASA/ESA {\em Hubble Space Telescope}
and 
the {\em Spitzer Space Telescope}.
The Space Telescope Science Institute (STScI) is operated by the Association of Universities for Research in Astronomy, 
Inc.~(AURA) under NASA contract NAS 5-26555. 
ACS was developed under NASA contract NAS 5-32864.
The Spitzer Space Telescope is operated by the Jet Propulsion Laboratory, California Institute of Technology under a contract with NASA.

This work utilizes gravitational lensing models produced by PIs Bradac, Ebeling, Zitrin \& Merten, Sharon, and Williams 
funded as part of the \HST\ Frontier Fields program conducted by STScI. 
The \HST\ image mosaics were produced by the Frontier Fields Science Data Products Team at STScI.
We obtained both the images and lens models from the Mikulski Archive for Space Telescopes (MAST).
And we thank these teams for all their efforts.

We also thank the STScI and SSC directors, 
their teams implementing the observing programs and reducing the data,
and especially the Frontier Fields PI Jennifer Lotz
for all of her hard work running this program.


}



{\it Facilities:}
\facility{HST (WFC3, ACS)};
\facility{Spitzer (IRAC)}

\bibliographystyle{astroads}
\bibliography{papersb}






\end{document}